\documentclass[review,3p]{elsarticle}
\usepackage{graphicx}
\usepackage{amsthm}
\usepackage{amssymb,color}
\usepackage{bm,amsmath}
\usepackage{subfigure}

\theoremstyle{remark}

\journal{Elseiver Science}

\begin{document}

\begin{frontmatter}



\title{Volume-averaged macroscopic equation for fluid flow in moving porous media}


\author[label1,label3]{Liang Wang}
\address[label1]{State Key Laboratory of Turbulence and Complex Systems, Peking University, Beijing, 100871, P.R. China}
\author[label2,label3]{Lian-Ping Wang}
\address[label2]{Department of Mechanical Engineering, University of Delaware, Newark, DE 19716-3140, USA}
\author[label3]{Zhaoli Guo\corref{cor1}}
\ead{zlguo@hust.edu.cn}
\address[label3]{State Key Laboratory of Coal Combustion, Huazhong University
of Science and Technology, Wuhan, 430074, P.R. China}
\cortext[cor1]{Corresponding author.}
\author[label1]{Jianchun Mi}

\begin{abstract}
Darcy's law and the Brinkman equation are two main models used for creeping fluid flows inside moving permeable particles. For these two models, the time derivative and the nonlinear convective terms of fluid velocity are neglected in the momentum equation. In this paper, a new momentum equation including these two terms are rigorously derived from the pore-scale microscopic equations by the volume-averaging method, which can reduces to Darcy's law and the Brinkman equation under creeping flow conditions. Using the lattice Boltzmann equation method, the macroscopic equations are solved for the problem of a porous circular cylinder moving along the centerline of a channel. Galilean invariance of the equations are investigated both with the intrinsic phase averaged velocity and the phase averaged velocity. The results demonstrate that the commonly used phase averaged velocity cannot serve as the superficial velocity, while the intrinsic phase averaged velocity should be chosen for porous particulate systems.

\end{abstract}


\end{frontmatter}


\section{Introduction}
\label{Sec:Intro}
The motion of permeable particles in a fluid has long received considerable attention in many fields such as colloid science, chemical, biomedical and environmental engineering. Because the fluid can penetrate into a permeable particle, there is a flow relative to the rigid skeleton of the porous medium. The hydrodynamic fields outside and inside the particles need to be treated together, which differs much from those of solid impermeable particles \cite{ChenSB99}. Numerous studies have been devoted to the understanding of transport phenomena in moving porous media for applications, such as sedimentation, agglomeration, flotation and filtration.
\par
In order to obtain the fluid velocity within a permeable particle, conservation equations which accurately govern the fluid flows are required for the permeable region \cite{Higdon81}. Under creeping flow conditions and considering the resistance force from the solid surface of moving porous media, two models for the fluid motion within porous media are commonly employed in the literature, i.e., Darcy's law and the Brinkman equation. Using the Stokes equation and Darcy's law, Payatakes and Dassios \cite{Payatakes87} investigated the motion of a porous sphere toward a solid planar wall. Later, Burganos et al. \cite{Burganos92} provided a revision to their work with respect to the drag force exerted on the permeable particle. Owing to the negligence of the viscous dissipation term, the momentum equation for the interior fluid of the porous media involves only first-order spatial derivatives in Darcy's law, while the momentum equation for the outside fluid includes spatial derivatives up to the second-order. This brings out a general fact that Darcy's law is confined to the case that the permeability of the porous media is sufficiently low. Meanwhile, the continuity in both the fluid velocity and the stress at the interface between the permeable medium and the exterior fluid are not guaranteed \cite{ChenSB99}. Complementary boundary treatment is hence needed to satisfy the continuity at the interface of a moving porous medium \cite{ChenSB98,Michalopoulou92,Michalopoulou93}. In contrast, in the Brinkman equation, the velocity-gradient term corresponding to viscous dissipation of the fluid within the porous medium is incorporated in the momentum equation, and the continuity of the fluid velocity is fulfilled at the surface of the porous body. From this point of view, the Brinkman equation is more suitable than Darcy's law in the porous particulate systems.
\par
Based on the Brinkman equation, numerous theoretical and computational studies have been conducted concerning moving permeable particles. For example, Jones \cite{Jones78a,Jones78b,Jones78c} calculated the forces and torques on moving porous particles with a method of reflection, and some studies were also carried out to investigate the suspension flow of porous particles using the Brinkman equation with other accompanying methods \cite{ChenSB99,Reuland78,Abade10a,Cichocki11}. Recently, the Brinkman model was also employed to study the hydrodynamic motion and interactions of composite particles \cite{ChenSB98,Ander&Kim87,Chen&Ye00,Abade12}.
\par
In all of the aforementioned studies, the flow inside a moving permeable particle is described using either Darcy's law or the Brinkman model. In these two models, the transient term and the nonlinear inertial term are not included in the momentum equation. It follows that there is no mechanism to treat the unsteady evolution of flow fields, and the flow Reynolds number for fluid flows must be kept sufficiently small. However, as noted by Wood \cite{Wood07}, the inertial effect on the flow and transport in porous media should be considered in many practical applications. To the best of our knowledge, no theoretical and numerical works have been developed to address this limitation. A new model is therefore desired for moving porous particulate systems at finite flow Reynolds numbers.
\par
On the other hand, due to the complexity of internal geometries and interfacial structures, it is impractical to solve the microscopic conservation equations inside the pores. A preferable approach \cite{Hassan&Gray79} is to average the microscopic equations inside porous particles over a representative elementary volume (REV), the size of which is assumed to be much larger than the characteristic size of pore structures but much smaller than the domain. Evidence from the literature indicates that a set of macroscopic equations at this scale can be derived through a rigorous volume-averaging procedure \cite{Anderson67,Whitaker67,Slattery67}, such as the solidification process of multicomponent mixtures \cite{Meck&Vis88,Ni&Bech91}, the flow through the interdendritic mushy zone \cite{Ganesan90}, the non-Newtonian fluid flows in porous media \cite{Getachew98}, and the flow in a stationary porous medium \cite{Hsu&Cheng90}. However, it appears that a general form of the volume-averaged or macroscopic momentum equation, where the transient as well as the nonlinear inertial terms are included, has not yet been developed for a moving porous medium.

The foregoing review of the literature has prompted us to derive more general governing equations for a moving porous medium, using the volume-averaging procedure. This is the main objective of the present work. Meanwhile, in addition to the phase average velocity commonly used for porous flows \cite{Ni&Bech91,Ganesan90,Hsu&Cheng90,Whitaker73}, the intrinsic phase average velocity is also employed in the literature \cite{Whitaker86,Yang13,Smit11}. For example, Yang et al. \cite{Yang13} recently used the intrinsic phase average form of the flow velocity in the macroscopic equations, while adopted the phase average form in the flow resistance term. Similar disparity in the fluid velocity is also found in the momentum equation used by Smit et al. \cite{Smit11}. In general, it is still not clear which kind of volume-averaged velocity should be used, especially for the case of moving porous media considered here. To our knowledge, no studies focused on this issue have been reported till now. This indicates the need for investigating the correct choice of superficial velocity from several possible volume-averaged velocities, which is another objective of this work.

In the following, the averaging theorems regarding the time derivatives and spatial derivatives are first presented. The macroscopic equations for the incompressible flow in a moving porous medium are then derived. To solve the derived macroscopic equations, a lattice Boltzmann equation (LBE) method \cite{Ddhum92,Lalleme&Luo00,Pan06} is employed, and numerical simulations in two frames of reference are carried out to investigate what kind of volume-averaged fluid velocity should be chosen. The results show that Galilean invariance of the macroscopic equations can be obtained only with the intrinsic phase averaged velocity, while the use of the phase averaged velocity will break the Galilean invariance.

\section{The method of volume averaging}\label{Sec:VAM}
In this work, the macroscopic governing equations for the fluid flow in a moving permeable body will be derived rigorously by averaging the microscopic continuity and momentum equations over a REV. To this end, the averaging theorems are needed to relate the average of the derivative to the derivative of the average. As shown in the literature \cite{Anderson67,Whitaker67,Slattery67,Whitaker73,Gray75,Howes85,Crapiste86,Gray13}, a number of authors developed these theorems forming the basis of the volume-averaging method. In this section, we will briefly review the invoked averaging theorems for subsequent derivations.

The flow in a moving porous medium is composed of fluid and solid phases. Assume that the fluid phase occupies a volume of $V_f$ in a representative volume $V$ within the porous medium. The volume occupied by the solid phase is hence $V_s=V-V_f$. In the study of multiphase transport process in porous media, it is important to distinguish three versions of volume averaged quantities \cite{Whitaker73}. The first of these is the \emph{intrinsic phase average} defined by
\begin{equation}
   \langle\psi_k\rangle^k=\frac{1}{V_k}\int_{V_k}\psi_k dV,
\end{equation}
where $V_k$ represents the volume of the $k$-phase within the representative volume $V$, $\psi_k$ is a quantity associated with the $k$-phase, and $k\in{\{f,s\}}$ with `f' and `s' respectively denoting the fluid and solid phases. The second is the \emph{phase average} which is the most commonly encountered averaged quantity \cite{Whitaker67,Slattery67}
\begin{equation}
   \langle\psi_k\rangle=\frac{1}{V}\int_{V_k}\psi_k dV
\end{equation}
With the two definitions, one can obtain the following relation
\begin{equation}\label{Eq:TwoRel}
   \langle\psi_k\rangle=\varepsilon_k\langle\psi_k\rangle^k,
\end{equation}
where $\varepsilon_k=V_k/V$ is the local volume fraction of the $k$-phase. The third average is the \emph{spatial average} of a quantity $\psi$ given by
\begin{equation}
   \langle\psi\rangle=\frac{1}{V}\int_{V}\psi dV,
\end{equation}
which assume $\psi$ can be defined in both phases and the average is taken over the fluid and solid phases. As noted by Whitaker \cite{Whitaker73}, the spatial average appears as an unimportant variable especially in the volume-averaged equations. Thus, our discussions on the superficial velocity will focus on the other two averages as defined above and consistently in this paper.

The macroscopic conservation equations are obtained by averaging the microscopic equations over a representative volume, and the following averaging theorems \cite{Whitaker73,Gray75} provide the basis to derive the macroscopic equations:
\begin{subequations}\label{Theo:average}
 \begin{equation}\label{Theorem1}
 \left\langle\frac{\partial \psi_k}{\partial t}\right\rangle=\frac{\partial\langle\psi_k\rangle}{\partial t}
    -\frac{1}{V}\int_{A_k}\psi_k\bm{w}_k\cdot\bm{n}_k dA,
 \end{equation}
 \begin{equation}\label{Theorem2}
   \langle\nabla\psi_k \rangle=\nabla\langle\psi_k\rangle+\frac{1}{V}\int_{A_k}\psi_k\bm{n}_k dA,
 \end{equation}
 \begin{equation}\label{Theorem3}
   \langle\nabla\cdot\bm{\psi_k}\rangle=\nabla\cdot\langle\bm{\psi_k}\rangle+\frac{1}{V}\int_{A_k}\bm{\psi_k}\cdot\bm{n}_k dA,
 \end{equation}
\end{subequations}
where $A_k$ is the interfacial area between the two phases within the averaging volume $V$, $\bm{w}_k$ represents the velocity of the interface $A_k$, $\bm{n}_k$ denotes the unit outwardly directed normal vector for the $k$-phase, and $\bm{\psi_k}$ is a vector quantity of the $k$-phase.

When $\psi_f=1$, we obtain from Eq. \eqref{Theorem2} the identity concerning the porosity $\varepsilon$, defined by $\varepsilon=V_f/V$, as follows
\begin{equation}\label{Eq:SpaEpison}
  \nabla\varepsilon=-\frac{1}{V}\int_{A_f}\bm{n}_f dA
\end{equation}
In addition, the following averaging theorem can be derived \cite{Ni&Bech91,Gray75}
\begin{equation}\label{Eq:Moditheo}
  \frac{1}{V}\int_{A_k}\langle\psi_k\rangle^k\bm{n}_k dA=-\langle\psi_k\rangle^k\nabla\varepsilon_k
\end{equation}

\section{Macroscopic equations}
We consider the motion of a porous body suspended in a Newtonian fluid. We assume that the REV scale is much smaller than the size of the body, but is much larger than the pore size. The porosity and permeability of the body are $\varepsilon$ and $K$, respectively. With the translational velocity $\bm{U}_p$ and rotational velocity $\bm{\Omega}_p$, the porous skeleton of the particle moves rigidly with velocity  $\bm{u}_s=\bm{U}_p+\bm{\Omega}_p\times\bm{r}$, where the position vector $\bm{r}$ is measured from the body center. At the surface of the permeable body, the microscopic fluid velocity and stress tensor are continuous. Since the porous media moves with a rigid-body motion, the translational and rotational velocities of the particle are unchanged even after taking the intrinsic phase average. In the following discussions, the interior of the porous particle is assumed to be homogeneous, and $\langle\bm{u}_s\rangle^s$ is denoted as $\bm{V}_p$.

Clearly, the fluid flow outside the porous particle is governed by the Navier-Stokes equations. As noted in the introduction, Darcy's law and the Brinkman equation are two main models employed in the literature to treat the flow at the REV scale inside the porous particle. Here we shall derive a more general REV model that includes the transient and nonlinear inertia terms by applying a rigorous volume averaging to the microscopic governing equations in the pores. In this section, we will present our analysis starting with the intrinsic phase averages, and then develop governing equations for alternative averages based on their inter-relations among different averaged quantities.

The volume averaging process starts with the equations that govern the microscopic fluid flow at the pore scale. The continuity and momentum equations are
\begin{equation}\label{Eq:MicMass}
   \frac{\partial\rho_f}{\partial t}+\nabla\cdot(\rho_f \bm{u}_f)=0,
\end{equation}
\begin{equation}\label{Eq:MicMomentum}
   \frac{\partial(\rho_f\bm{u}_f)}{\partial t}+\nabla\cdot(\rho_f \bm{u}_f\bm{u}_f)=-\nabla p_f+\nabla\cdot\bm{\tau}_f+\rho_f\bm{g},
\end{equation}
where $\rho_f$ is the fluid density, $\bm{u}_f$ is the fluid velocity, $p_f$ is the pressure, $\bm{g}$ is the acceleration of external body force, $\bm{\tau}_f$ is the stress tensor given by
$\bm{\tau}_f=\mu\left(\nabla\bm{u}_f+(\nabla\bm{u}_f)^T\right)$ with $\mu$ being the fluid dynamic viscosity, and the superscript $T$ denotes the transpose operator.

Applying the theorems \eqref{Theorem1} and \eqref{Theorem3} to the left hand side of Eq. \eqref{Eq:MicMass}, and assuming the fluid density to be uniform in the integral volume $V_f$ ( i.e., incompressible flow), we obtain
\begin{equation}\label{Eq:Masstemp}
   \frac{\partial \left(\varepsilon\rho_f\right)}{\partial t}+\nabla\cdot\bigl(\rho_f\langle\bm{u}_f\rangle\bigr)
      =-\frac{1}{V}\int_{A_f}\rho_f(\bm{u}_f-\bm{w}_f)\cdot\bm{n}_fdA.
\end{equation}

In the above equation, $\rho_f(\bm{u}_f-\bm{w}_f)\cdot\bm{n}_f$ represents the mass flux across the interface from the fluid phase to the solid phase. Since no phase change occurs in the considered system, the area integral of this term will be zero. Thus, the macroscopic continuity equation can be written in the form of the intrinsic average as
\begin{equation}\label{Eq:MacroMass}
   \frac{\partial \left(\varepsilon\rho_f\right)}{\partial t}
    +\nabla\cdot\bigl(\varepsilon\rho_f\langle\bm{u}_f\rangle^f\bigr)=0.
\end{equation}

Next turn to the momentum equation. Volume-averaging of Eq. \eqref{Eq:MicMomentum} with the aid of Eq. \eqref{Theo:average} leads to
\begin{align}\label{Eq:MoPhaseAv}
  \frac{\partial \left(\rho_f\langle\bm{u}_f\rangle\right)}{\partial t}+\nabla\cdot\left(\rho_f\langle\bm{u}_f\bm{u}_f\rangle\right)
  +\frac{1}{V}\int_{A_f}\rho_f\bm{u}_f\left(\bm{u}_f-\bm{w}_f\right)\cdot\bm{n}_f dA \notag\\
  =-\nabla\langle p_f\rangle+\nabla\cdot\langle\bm{\tau}_f\rangle+\frac{1}{V}\int_{A_f}\left(-p_f\bm{I}+\bm{\tau}_f\right)\cdot\bm{n}_f dA+\rho_f\langle\bm{g}\rangle,
\end{align}
where $\bm{I}$ is the identity tensor. The third term in Eq. \eqref{Eq:MoPhaseAv} represents the momentum exchange due to the interphase mass transfer. Because of the no-slip condition at the interfacial surface and the assumption of no phase change, this term is zero as $\bm{u}_f=\bm{w}_f$.

To obtain the volume-averaged equation for the motion of fluid, the term involving the average of product in Eq. \eqref{Eq:MoPhaseAv}, $\langle\bm{u}_f\bm{u}_f\rangle$, needs to be changed to the product of averages. For this purpose, we follow the spatial decomposition suggested by Gray \cite{Gray75} and Hsu and Cheng \cite{Hsu&Cheng90}, and define the microscopic velocity $\bm{u}_f$ by
\begin{equation}\label{Eq:Decom}
  \bm{u}_f=\langle\bm{u}_f\rangle^f+\bm{u}_f',
\end{equation}
where $\langle\bm{u}_f\rangle^f$ and $\bm{u}_f'$ are the intrinsic phase average and the fluctuating component, respectively. Note that $\langle\bm{u}_f'\rangle^f=0$ provided $\langle\bm{u}_f\rangle^f$ is well defined. Then,
\begin{equation}\label{Eq:dispersion}
  \langle\bm{u}_f\bm{u}_f\rangle=\varepsilon\langle\bm{u}_f\rangle^f\langle\bm{u}_f\rangle^f
      +\varepsilon\langle\bm{u}_f'\bm{u}_f'\rangle^f,
\end{equation}
where the second term on the right-hand-side represents the average dispersion vector \cite{Whitaker67,Hsu&Cheng90,Gray75}. As $\varepsilon\rightarrow 1$ or $\varepsilon\rightarrow 0$, the contribution of this term to the momentum balance can be ignored. Because of the tortuous structure resided in the pore spaces and the no-slip condition at the solid-fluid interface, this dispersive term may contribute to the momentum transfer for some special ranges of $\varepsilon$ \cite{Ganesan90}. Nonetheless, same as that in Ref. \cite{Ganesan90}, we will neglect this term in the present work. Actually, Hsu and Cheng \cite{Hsu&Cheng90} assume that $\bm{u}_f' \ll \langle\bm{u}_f\rangle^f $, then the same term in Eq. \eqref{Eq:dispersion} is of higher order and is neglected.

Thus, Eq. \eqref{Eq:MoPhaseAv} can also be rewritten in terms of the intrinsic phase average velocity
\begin{align}\label{Eq:MoIntPhaseAv}
  &\frac{\partial\left(\rho_f\varepsilon\langle\bm{u}_f\rangle^f\right)}{\partial t}+\nabla\cdot\left(\rho_f\varepsilon\langle\bm{u}_f\rangle^f\langle\bm{u}_f\rangle^f\right)\notag\\
  \;\;\;&=-\nabla\left(\varepsilon\langle p_f\rangle^f\right)+\nabla\cdot\left(\varepsilon\langle\bm{\tau}_f\rangle^f\right)
  +\frac{1}{V}\int_{A_f}\left(-p_f\bm{I}+\bm{\tau}_f\right)\cdot\bm{n}_f dA+\varepsilon\rho_f\bm{g},
\end{align}
where the acceleration of $\bm{g}$ is assumed to be constant over the averaging volume.
Denoting $\bm{\sigma}_f=-p_f\bm{I}+\bm{\tau}_f$ and following the treatment given in Refs. \cite{Pros&Tryg09,Harlow75} for the first three terms on the right-hand-side of Eq. \eqref{Eq:MoIntPhaseAv}, the following identity holds
\begin{equation}\label{Eq:Transint}
  \nabla\cdot\left(\varepsilon\langle\bm{\sigma}_f\rangle^f\right)
  +\frac{1}{V}\int_{A_f}\bm{\sigma}_f\cdot\bm{n}_f dA=\varepsilon\nabla\cdot\left(\langle\bm{\sigma}_f\rangle^f\right)
  +\frac{1}{V}\int_{A_f}\left(\bm{\sigma}_f-\langle\bm{\sigma}_f\rangle^f\right)\cdot\bm{n}_f dA,
\end{equation}
in which Eq. \eqref{Eq:SpaEpison} is incorporated into the derivation. The interfacial integral on the right-hand side of Eq. \eqref{Eq:Transint} pertains to the total average drag force due to the interactions between fluid and moving solid, which is denoted as
\begin{equation}\label{Eq:Fordenote}
  \bm{B}=\frac{1}{V}\int_{A_f}\left(\bm{\sigma}_f-\langle\bm{\sigma}_f\rangle^f\right)\cdot\bm{n}_f dA
\end{equation}

Now we seek the expression with appropriate macroscopic variables for the interphase force $\bm{B}$. It is noted in the literature \cite{Whitaker67,Ni&Bech91,Ganesan90,Whitaker73,Gray75} that the force $\bm{B}$ depends on the relative intrinsic average velocity of the solid and fluid. Several empirical expressions are available in the literature for this interfacial force. For a static porous medium, Hsu and Cheng \cite{Hsu&Cheng90} and Getachew et al. \cite{Getachew98} performed a dimensional analysis respectively for Newtonian and non-Newtonian flows to represent the interfacial force. In the context of the relative motion of fluid and solid in the solidification system, Ni and Beckermann \cite{Ni&Bech91} suggested an expression of the fluid-solid interaction forces in analogy with Darcy's law, and Ganesan and Poirier \cite{Ganesan90} considered a nonlinear resistance term which is expressed as the square of the relative velocity. These previous studies suggest that the interphase force $\bm{B}$ can be modeled by the relative velocity between the solid and fluid. In this paper, we follow the work of Hsu and Cheng \cite{Hsu&Cheng90} but account for the movement of porous medium to derive the expression of $\bm{B}$. Using the Ergun equation, we write
\begin{equation}
  \bm{B}=-\varepsilon\left[\frac{\mu\varepsilon}{K}\left(\langle\bm{u}_f\rangle^f-\bm{V}_p\right)
   +\rho_f\frac{\varepsilon^2F_\varepsilon}{\sqrt{K}}\left(\langle\bm{u}_f\rangle^f
   -\bm{V}_p\right)\left|\langle\bm{u}_f\rangle^f-\bm{V}_p\right|\right],
\end{equation}
where the permeability $K$ and geometric function $F_\varepsilon$ of the porous medium obey the Ergun's expression
\begin{equation}
  K=\frac{\varepsilon^3d_p^2}{150(1-\varepsilon)^2}, \quad F_\varepsilon=\frac{1.75}{\sqrt{150\varepsilon^3}},
\end{equation}
where $d_p$ is the diameter of grain particles within the porous solid particle. It is noted that the relation of $\langle\bm{u}_s\rangle^s=\bm{V}_p$ is employed in the derivation.

Using Eq. \eqref{Theorem2} and recognizing that the viscosity and the porosity are constant in the averaging volume, we have
\begin{equation}\label{Eq:Stress}
   \varepsilon\langle\bm{\tau}_f\rangle^f =\mu\biggl[\nabla\left(\varepsilon\langle\bm{u}_f\rangle^f\right)
      +\left[\nabla\left(\varepsilon\langle\bm{u}_f\rangle^f\right)\right]^T +\frac{1}{V}\int_{A_f}\bigl(\bm{u}_f\bm{n}_f+ \bm{n}_f\bm{u}_f\bigr)dA\biggr]
 \end{equation}
For the surface integral in Eq. \eqref{Eq:Stress}, the no-slip boundary condition at the fluid-solid interface ensures that the interfacial fluid velocity can be replaced by the solid velocity. Similar to Eq. \eqref{Eq:Decom}, the solid velocity is first decomposed as $\bm{u}_s=\langle\bm{u}_s\rangle^s+\bm{u}_s'$. Noticed that $\bm{u}_s'$ represents the deviation of the microscopic velocity of solid from its intrinsic phase average velocity. Meanwhile, the skeleton of a porous particle behaves as a rigid body motion in which the translational and rotational velocity at each point in the solid region are the same. This suggests that the interfacial average of $\bm{u}_s$ is approximately equal to $\langle\bm{u}_s\rangle^s$, that is, the surface integral of $\bm{u}_s'$ at the interface of solid and fluid is very small. With this approximation, Eq. \eqref{Eq:Stress} and combining Eq. \eqref{Eq:Moditheo}, the averaged shear stress takes the form:
\begin{equation}\label{Eq:StressModi}
   \varepsilon\langle\bm{\tau}_f\rangle^f =\mu\biggl[\nabla\left(\varepsilon\langle\bm{u}_f\rangle^f\right)
      +\left[\nabla\left(\varepsilon\langle\bm{u}_f\rangle^f\right)\right]^T-
      \langle\bm{u}_s\rangle^s\nabla\varepsilon-\nabla\varepsilon\langle\bm{u}_s\rangle^s\biggr]
 \end{equation}
As noted previously, the porosity is uniform over the solid region. Consequently, the last terms on the right-hand side of Eq. \eqref{Eq:StressModi} will vanish, and the final expression for the macroscopic shear stress is
\begin{equation}\label{Eq:StressTerm}
   \varepsilon\langle\bm{\tau}_f\rangle^f =\mu\varepsilon\biggl[\nabla\left(\langle\bm{u}_f\rangle^f\right)
      +\left[\nabla\left(\langle\bm{u}_f\rangle^f\right)\right]^T\biggr]
 \end{equation}

Putting all the terms together, the macroscopic momentum equation for the fluid flows inside moving porous particle can be written in terms of the intrinsic phase averaged variables as
\begin{align}\label{Eq:IntrPhaseAvR}
   &\frac{\partial\left(\rho_f\varepsilon\langle\bm{u}_f\rangle^f\right)}{\partial t}+\nabla\cdot\left(\rho_f\varepsilon\langle\bm{u}_f\rangle^f\langle\bm{u}_f\rangle^f\right)\notag\\
     &\qquad\qquad=-\varepsilon\nabla\left(\langle p_f\rangle^f\right)+\mu\varepsilon\nabla\cdot\biggl[\nabla\left(\langle\bm{u}_f\rangle^f\right)
      +\left[\nabla\left(\langle\bm{u}_f\rangle^f\right)\right]^T\biggr]+\bm{F},
\end{align}
where $\bm{F}$ is the total body force including the resistance from the porous medium and other external force
\begin{equation}\label{IntrPhaForce}
  \bm{F}=-\varepsilon\left[\frac{\mu\varepsilon}{K}\left(\langle\bm{u}_f\rangle^f-\bm{V}_p\right)
   +\rho_f\frac{\varepsilon^2F_\varepsilon}{\sqrt{K}}\left(\langle\bm{u}_f\rangle^f
   -\bm{V}_p\right)\left|\langle\bm{u}_f\rangle^f-\bm{V}_p\right|\right]+\varepsilon\rho_f\bm{g}.
\end{equation}

Eqs. \eqref{Eq:MacroMass}, \eqref{Eq:IntrPhaseAvR} and \eqref{IntrPhaForce} are the volume-averaged equations written using the intrinsic phase-averaged quantities. Generally, the phase average velocity is used as the superficial velocity for flows in porous media. At the same time, as noted in the introduction, the intrinsic phase average is also employed in the literature. However, no studies have been conducted to identify the superficial velocity from the volume-averaged velocities when the porous medium moves. In this paper, Galilean invariance of the volume-averaged equations will be numerically investigated, and we use the intrinsic phase average velocity as an alternative to the phase average velocity when performing numerical validations. It will be shown later in the numerical experiments that Galilean invariance can be obtained when using the intrinsic phase average velocity, while the results using the phase average velocity are not Galilean invariant. The macroscopic equations expressed in terms of the intrinsic phase average velocity will be given next.
\\[2.5mm]
\emph{Macroscopic equations in terms of the intrinsic phase average velocity}
\par
Application of Eq. \eqref{Eq:MacroMass} to the two terms on the left-hand-side of Eq. \eqref{Eq:IntrPhaseAvR} provides
\begin{equation}
   \frac{\partial\left(\rho_f\varepsilon\langle\bm{u}_f\rangle^f\right)}{\partial t}+\nabla\cdot\left(\rho_f\varepsilon\langle\bm{u}_f\rangle^f\langle\bm{u}_f\rangle^f\right)
   =\varepsilon\rho_f\left[\frac{\partial\langle\bm{u}_f\rangle^f}{\partial t}+\langle\bm{u}_f\rangle^f\cdot\nabla\langle\bm{u}_f\rangle^f\right]
\end{equation}
Thus, based on Eqs. \eqref{Eq:MacroMass}, \eqref{Eq:IntrPhaseAvR} and \eqref{IntrPhaForce}, the following incompressible macroscopic equations can be obtained
\begin{subequations}\label{Eq:InPhavGove}
 \begin{equation}\label{Eq:MacroMassInPhav}
   \nabla\cdot \langle\bm{u}_f\rangle^f =0,
 \end{equation}
 \begin{align}\label{Eq:InPhavMoment}
   \rho_f\left[\frac{\partial\langle\bm{u}_f\rangle^f}{\partial t} +\langle\bm{u}_f\rangle^f\cdot\nabla\langle\bm{u}_f\rangle^f\right]
    =-\nabla\langle p_f\rangle^f +\mu\nabla^2\langle\bm{u}_f\rangle^f+\bm{F},
\end{align}
where
\begin{equation}\label{Eq:InPhavForce}
  \bm{F}=-\frac{\mu\varepsilon}{K}\left(\langle\bm{u}_f\rangle^f-\bm{V}_p\right)
   -\rho_f\frac{\varepsilon^2F_\varepsilon}{\sqrt{K}}\left(\langle\bm{u}_f\rangle^f
   -\bm{V}_p\right)\left|\langle\bm{u}_f\rangle^f-\bm{V}_p\right|+\rho_f\bm{g}.
\end{equation}
\end{subequations}

In order to illustrate the validity of the derived macroscopic equations with the intrinsic phase average velocity, the relevant dimensionless velocities are identified for comparison in the numerical experiments. By introducing characteristic quantities into Eq. \eqref{Eq:InPhavGove} and conducting non-dimensionalization procedure, the dimensionless versions of the macroscopic equations can be obtained. Corresponding to Eq. \eqref{Eq:InPhavGove}, the resulting dimensionless equations are
\begin{subequations}\label{Eq:InPhavGoveDimless}
 \begin{equation}\label{Eq:MacroMassInPhavDimless}
   \nabla\cdot \langle\bm{u}_f\rangle^f =0,
 \end{equation}
 \begin{align}\label{Eq:InPhavMomentDimless}
  \frac{\partial\langle\bm{u}_f\rangle^f}{\partial t} +\langle\bm{u}_f\rangle^f\cdot\nabla\langle\bm{u}_f\rangle^f
    =-\nabla \langle p_f\rangle^f +\frac{1}{Re}\nabla^2\langle\bm{u}_f\rangle^f+\bm{F},
\end{align}
where
\begin{equation}\label{Eq:InPhavForceDimless}
  \bm{F}=- \frac{\varepsilon}{ReDa}\left(\langle\bm{u}_f\rangle^f- \bm{V}_p\right)
   -\frac{\varepsilon^2 F_\varepsilon}{\sqrt{Da}}\left(\langle\bm{u}_f\rangle^f-\bm{V}_p\right)
   \left|\langle\bm{u}_f\rangle^f-\bm{V}_p\right|+\bm{g}.
\end{equation}
\end{subequations}

It is noteworthy that the flow in a moving porous medium is characterized by the porosity $\varepsilon$ and two dimensionless parameters from the above dimensionless equations. They are the Reynolds number $Re$ and Darcy number $Da$, respectively defined as
\begin{equation}
    Re=\frac{LU}{\nu},\qquad Da=\frac{K}{L^2},
\end{equation}
where $L$ and $U$ are the characteristic length and velocity, and $\nu$ is the fluid kinematic viscosity relating with the dynamic viscosity by $\mu=\rho_f\nu$.
\par
For the phase average form of the volume-averaged equations, it is necessary to note that still based on Eqs. \eqref{Eq:MacroMass} and \eqref{Eq:IntrPhaseAvR} and \eqref{IntrPhaForce}, the incompressible macroscopic equations expressed in terms of the phase average velocity can be directly derived (See \ref{AppSpa} for details). In addition, the numerical results with the phase average velocity will be used for comparison with the intrinsic phase average velocity.

The volume-averaged conservation equations derived above represent a more generalized model than Darcy's law and the Brinkman equation. In other words, Darcy's law and the Brinkman equation would be recovered when the fluid motion is sufficiently slow. This aspect is discussed next using the macroscopic equations in terms of the intrinsic phase average velocity.
\\[2.5mm]
\emph{Darcy's law}
\par
When the flow velocity is very low and the flow is steady, the transient term, the inertial term and the quadratic drag force term can be neglected in the momentum equation. These hold true for Eqs. \eqref{Eq:InPhavMoment} and \eqref{Eq:InPhavForce} in terms of intrinsic phase average velocity. In addition, the viscous term is assumed to be negligibly small.

Noting that $\langle\bm{u}_f\rangle=\varepsilon\langle\bm{u}_f\rangle^f$, and with the presence of external force, the momentum equation of Eq. \eqref{Eq:InPhavGove} at $\bm{V}_p=0$, reduces to
\begin{equation}\label{Eq:Darcy}
  \langle\bm{u}_f\rangle=-\frac{K}{\mu}\left[\nabla\langle p_f\rangle^f-\rho_f\bm{g}\right],
\end{equation}
which recovers the well-known Darcy's law. It is clear that in this model the pressure is the intrinsic phase average pressure, and the superficial velocity is the phase average one.
\\[2.5mm]
\emph{Brinkman equation}
\par
As mentioned in the introduction, the Brinkman equation is a modification of Darcy's law. Under the low-speed flow condition, the inertia term is dropped but the viscous dissipation term is retained in the Brinkman equation. Under similar assumptions, Eqs. \eqref{Eq:InPhavMoment} and \eqref{Eq:InPhavForce}(without the external force) yield
\begin{equation}\label{Eq:Brinkman}
  -\nabla\langle p_f\rangle^f+\mu \nabla^2\langle\bm{u}_f\rangle^f-\frac{\mu_e}{K}\left(\langle\bm{u}_f\rangle^f-\bm{V}_p\right)=0,
\end{equation}
where $\mu_e=\mu\varepsilon$. This equation has the same form as the Brinkman equation, and similar formula has been used in the literature \cite{ChenSB99,ChenSB98,Abade10a,Abade12}. In addition, we would like to point out that an equation of the same form as the Brinkman equation cannot be extracted from the derived volume-averaged equations with the phase average velocity. This result indicates that the macroscopic equations expressed with the intrinsic phase average velocity would be more preferable.
\par
In summary, we have presented both the macroscopic conservation equations and their dimensionless forms for the fluid flows inside a moving porous medium. In the limiting case as $\varepsilon=1$, i.e., in the absence of porous media, one can see that the value of $K$ or $Da$ will become infinite, and Eq. \eqref{Eq:InPhavGove} or Eq. \eqref{Eq:InPhavGoveDimless} reduces to the Navier-Stokes equations for pure fluid flows. While as $\varepsilon\rightarrow0$, i.e., for a moving solid particle, $K$ (or $Da$) approaches to zero, and thus the inner ``fluid" has the same velocity as the particle. Therefore, in the solution of the derived macroscopic equations, we can use different values of $\varepsilon$ in different regions so that the fluid flows can be calculated throughout the whole domain. Finally, when $\bm{V}_p=0$ and expressed with the phase average velocity, Eqs. \eqref{Eq:MacroMass}, \eqref{Eq:IntrPhaseAvR} and \eqref{IntrPhaForce} reduces to the macroscopic equations derived by Hsu and Cheng \cite{Hsu&Cheng90} for the flow inside a stationary porous medium.

\section{Numerical experiments}
\label{Sec:Numer}
In this section, in order to investigate the validity of the intrinsic phase average velocity as the superficial velocity for the derived macroscopic equations, numerical experiments will be performed on the flow in a moving porous medium. The dimensionless equations, Eq. \eqref{Eq:InPhavGoveDimless} are employed as the governing equations, and the LBE model is used as the numerical solver \cite{Guo&Shu13}. As noted previously, the macroscopic equations are valid for the flows both inside and outside the porous medium. Hence, the numerical solution of the equations can be carried out on a uniform grid system. For comparison, the other set of macroscopic equations with the phase average velocity, Eq. \eqref{Eq:PhavGoveDimless}, are also solved.

\begin{figure}
  \centering \includegraphics[width=4.8in]{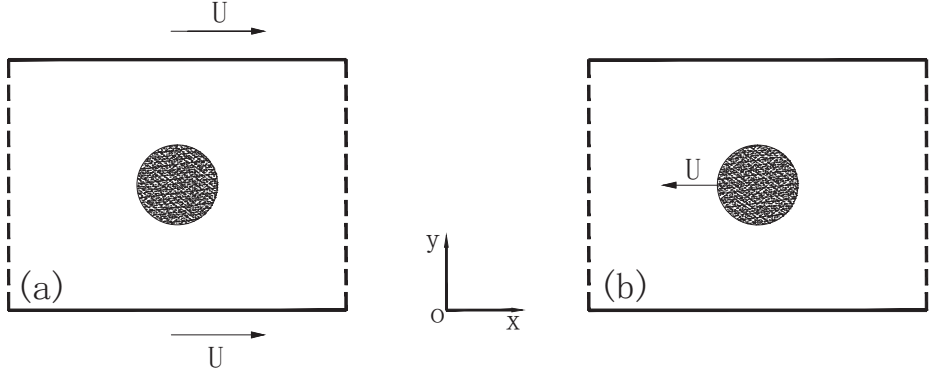}
  \caption{Configuration of the symmetric motion of a porous cylinder along a channel. Two frames of reference are performed: (a) the coordinate system is fixed on the porous cylinder. (b) the coordinate system is fixed on the flat wall.}
  \label{porsquaresch} \vspace{-4.0mm}
\end{figure}

The test problem is the flow past a porous cylinder. The geometric configuration is depicted in Fig. \ref{porsquaresch}, in which the porous cylinder moves along the channel centerline with a constant horizontal velocity $U$.
The simulations can be performed in two frames of reference, that is, the coordinate system is fixed on the cylinder ( Fig. \ref{porsquaresch} (a) ) or on the flat wall of the channel (Fig. \ref{porsquaresch} (b) ). For the sake of description, these two cases are denoted by R1 and R2, respectively. Galilean invariance of the macroscopic equations ensures that the relative motion between the porous cylinder and the flow in the channel is identical in the R1 and R2 frames. Therefore, by directly comparing the results calculated with the intrinsic phase average velocity and/or the phase average velocity in these two cases, we can ascertain which of the two volume-averaged velocities should be used as the superficial velocity.
\par
As for Fig. \ref{porsquaresch} (a), the porous cylinder is initially at rest. Both the upper and the lower walls move in $x$-direction with an uniform velocity $U$. As for Fig. \ref{porsquaresch} (b), the cylinder is moving with $U$ along the opposite $x$-direction, while the upper and lower walls of the channels are kept at fixed. In the two cases, the periodic boundary conditions are used in $x$-direction, and the half-way bounce-back scheme is applied to the channel side walls for the no-slip boundary conditions. The diameter of the cylinder is $D=0.24$, and the computational domain for the channel is $L\times H =15D\times 15D$, which is covered by a $360\times359$ square mesh. Initially, the center of the porous cylinder is located at $(x_0,y_0)=(7.5D,7.5D)$. The Reynolds number, defined by $Re=UD/\nu$, is 100.
\par
The numerical calculations are conducted using the LBE model ( see \ref{Appen:MRTMod} for details ). The relaxation rates in the LBE model are set as: $s_0=s_3=s_5=0$, $s_1=1.1$, $s_2=1.25$, $s_7=s_8=1/\tau$, while $s_4=s_6=8(2-s_7)/(8-s_7)$. Here, the value of $\tau$ is determined based on $1/Re=c_s^2(\tau-1/2)\delta_t$, and $c_s^2=1/3$, $\delta_t=\delta_x=1/100$. In order to keep the porous cylinder in the computational domain, a moving computational domain is used for the case of R2 during the simulations. As recommended in the LBE method for force evaluation \cite{Mei&Luo02,Yu&Luo03}, the momentum-exchange method is used here to compute the hydrodynamic force on the porous cylinder.
\par
\begin{figure}
\begin{tabular}{cc}
\includegraphics[width=0.49\textwidth,height=0.3\textheight]{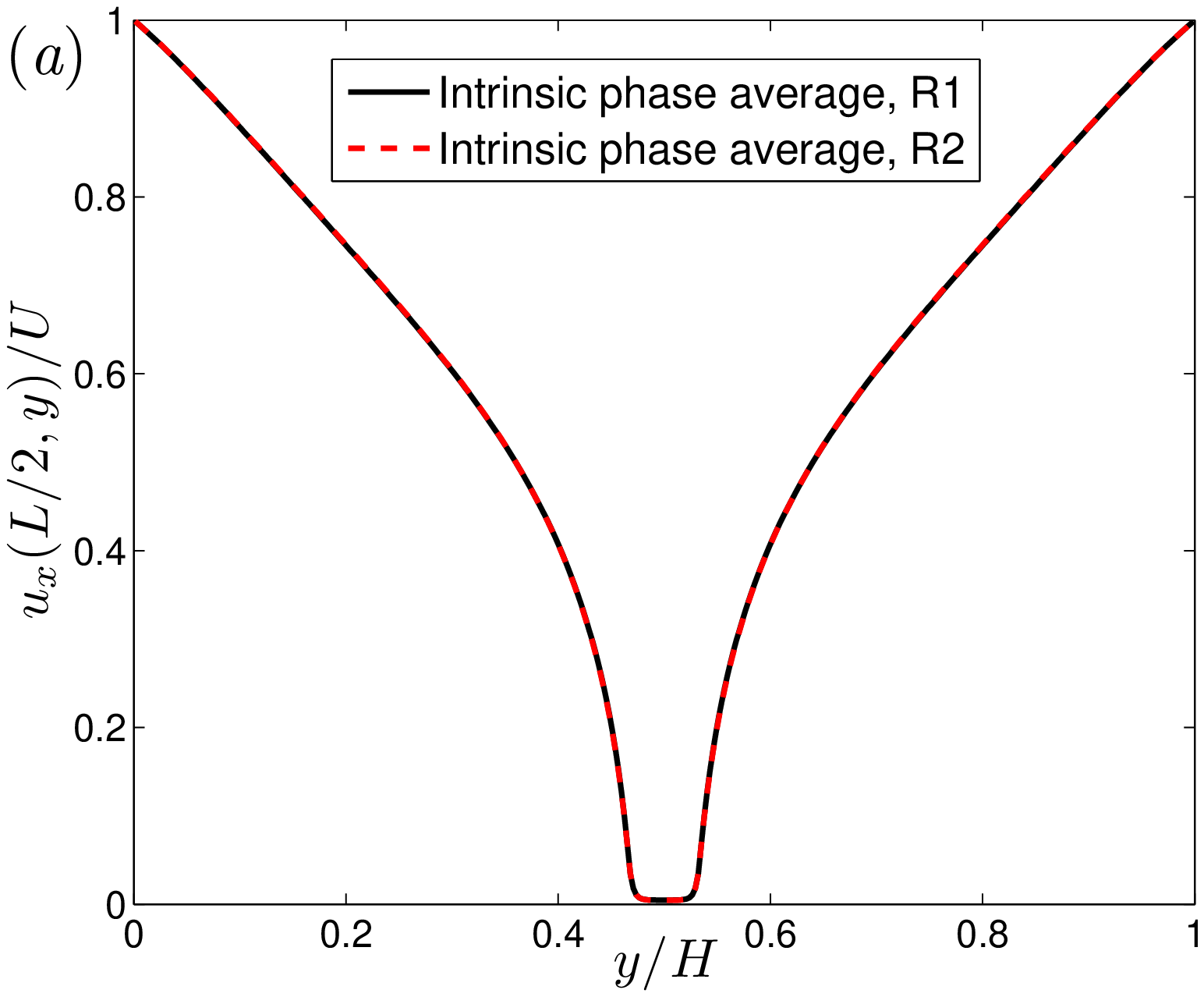}&
\includegraphics[width=0.49\textwidth,height=0.3\textheight]{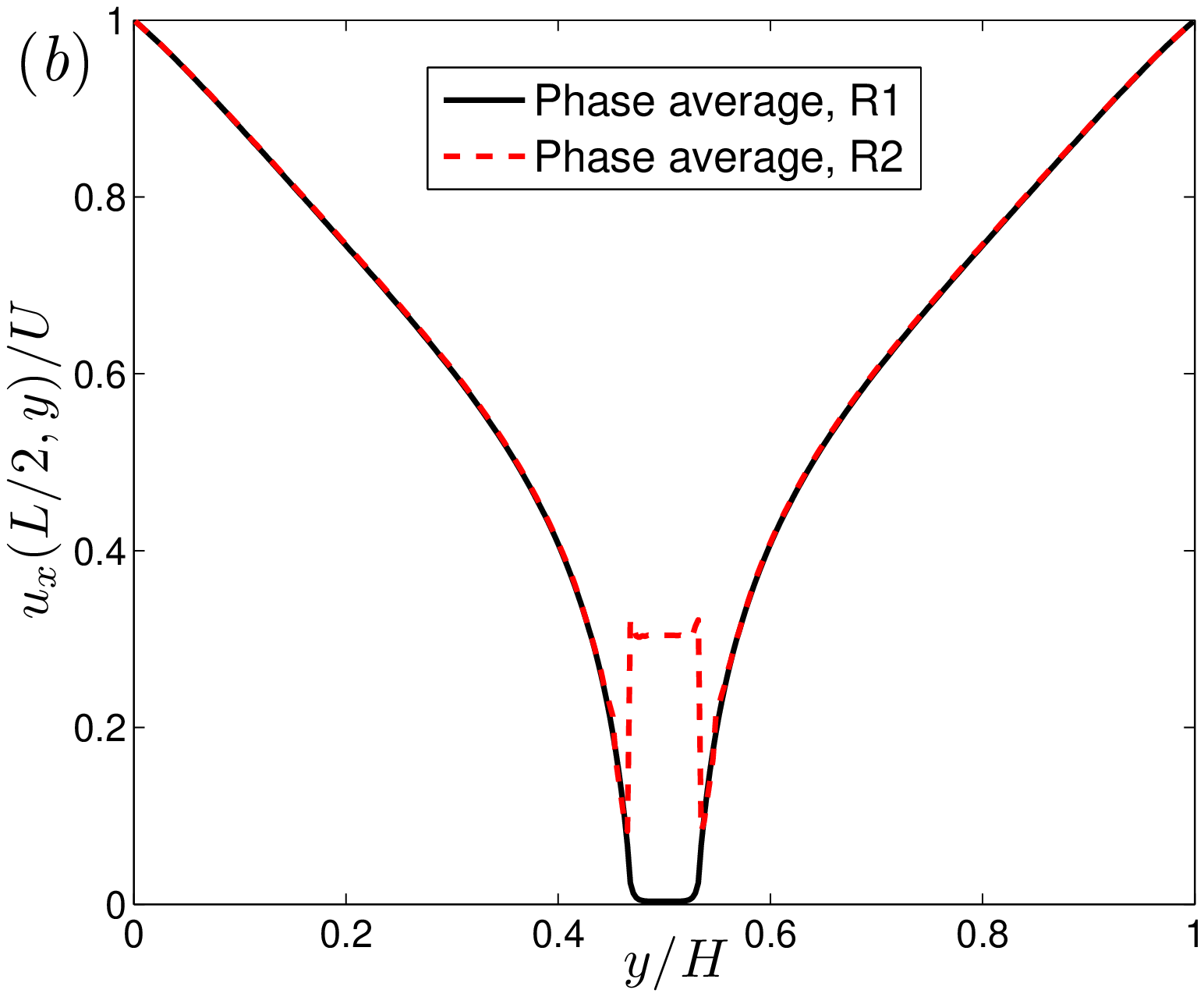}\\
\end{tabular}
\caption{Profiles of (a) the intrinsic phase average velocity and (b) the phase average velocity through the channel center at $\varepsilon=0.7$ and $Re=100$. The results are obtained using the two frames of reference. }
\label{Galifig1}
\end{figure}
\begin{figure}
\begin{tabular}{cc}
\includegraphics[width=0.49\textwidth,height=0.3\textheight]{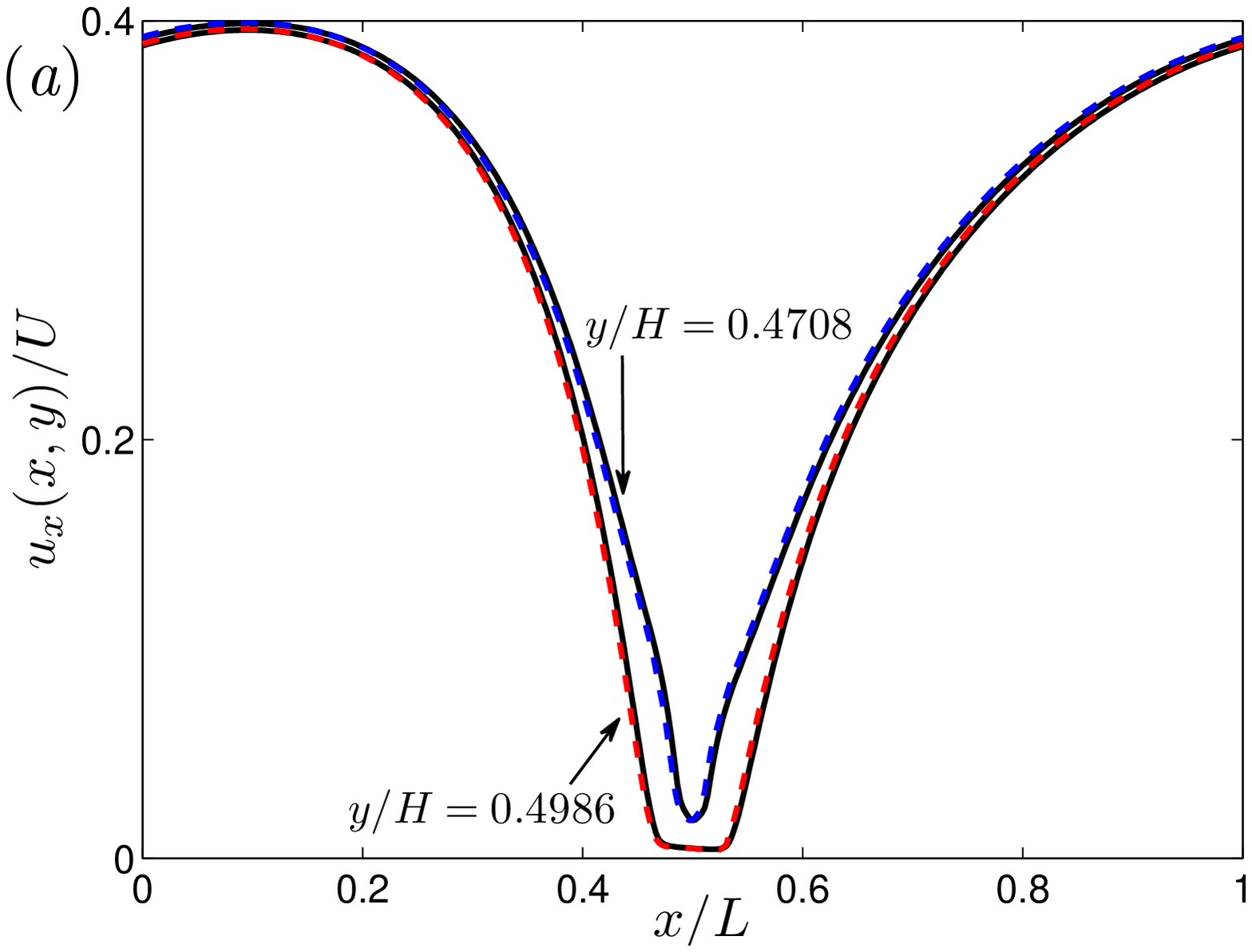}&
\includegraphics[width=0.49\textwidth,height=0.3\textheight]{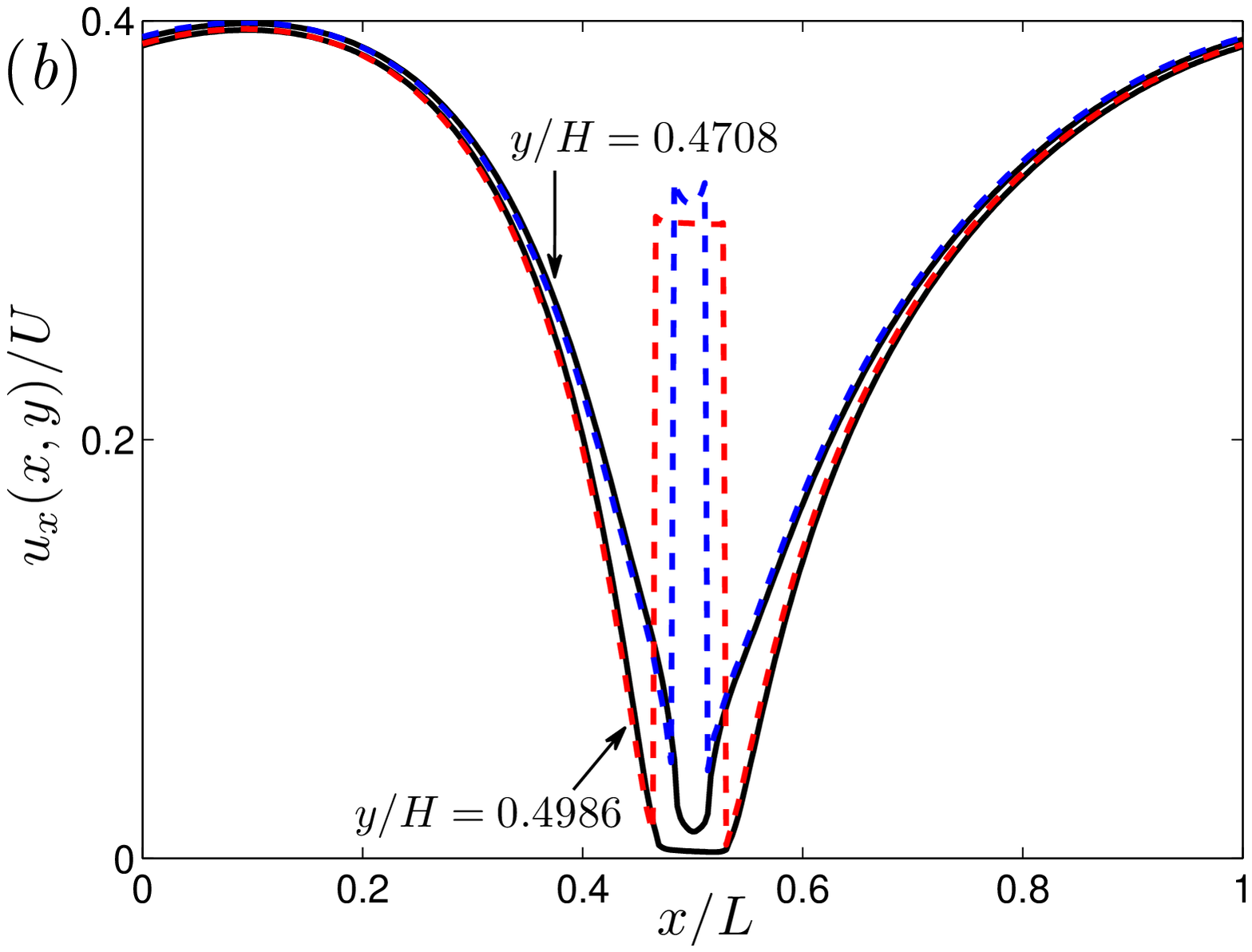}\\
\end{tabular}
\caption{Velocity profiles in the porous region along horizontal lines at $\varepsilon=0.7$. (\textit{a}) The intrinsic phase average velocity. (\textit{b}) The phase average velocity. Solid lines are the results in the frame of R1, and dashed lines are the results in the frame of R2.}
\label{fig:Velcomp1}
\end{figure}

In Fig. \ref{Galifig1}, the profiles of the intrinsic phase average velocity through the channel center are plotted for each frame of reference at $\varepsilon=0.7$. The profiles of the phase-averaged velocity are also shown in the figure.
As can be seen from the figure, in the pure fluid region, the results for both the intrinsic phase average velocity and the phase average velocity coincide with each other in the frame of R1 (i.e., $U=0$) and in the frame of R2 (i.e., $U\neq0$). However, in the porous region, it is not the case for the two average velocities in the two frames of reference. For the intrinsic phase average velocity, perfect agreement in the frames of R1 and R2 are still obtained, while for the phase average velocity an obvious deviation between the two frames are observed. Similar difference in the porous region can also be found between the two volume-averaged velocities along horizontal lines, which are shown in Fig. \ref{fig:Velcomp1}.
As displayed in the figure, the calculated intrinsic phase average velocities along horizontal lines are in good agreement in the frames of R1 and R2. In contrast to this, the phase average velocities disagree with each other in the two frames. The results from Figs. \ref{Galifig1} and \ref{fig:Velcomp1} clearly demonstrate that only the derived macroscopic equations with the intrinsic phase average velocity are Galilean invariant.

Furthermore, the difference between the frames of R1 and R2 in Fig. \ref{Galifig1} is measured in terms of the phase average velocity. It is found that the difference in the porous region between the two cases is $0.3$, which stems from the magnitude of $(1-\varepsilon)U$. This can be explained by the following analysis: Assume the intrinsic phase average velocity inside the porous cylinder is $u_0$ in the frame of R1, and thus the phase average velocity inside the cylinder is $\varepsilon u_0$ as the cylinder is fixed. In the particle moving frame of reference with velocity of $-U$, the intrinsic phase average velocity is Galilean invariant, and thus the intrinsic phase average velocity inside the porous cylinder is now $u_0-U$. That is, the phase average velocity is $\varepsilon(u_0-U)$ inside the porous cylinder in the frame of R2. As a consequence, the phase average velocity relative to the cylinder is $\varepsilon(u_0-U)-(-U)=\varepsilon u_0+(1-\varepsilon)U$. This indicates that the phase average velocities in the two cases have the difference of $(1-\varepsilon)U$, which is exactly what is observed in Fig. \ref{Galifig1}. Therefore, the phase average velocity is indeed not Galilean invariant at the face value, and it should not be used as the superficial velocity in the derived macroscopic equations.

The hydrodynamic force on the porous cylinder as a function of time is also measured for the two frames. Considering that the lift force component is considerably small, we only compute the drag force component $F_x$ with the two volume-averaged velocities in the frames of R1 and R2. \begin{figure}
\begin{tabular}{cc}
\includegraphics[width=0.49\textwidth,height=0.308\textheight]{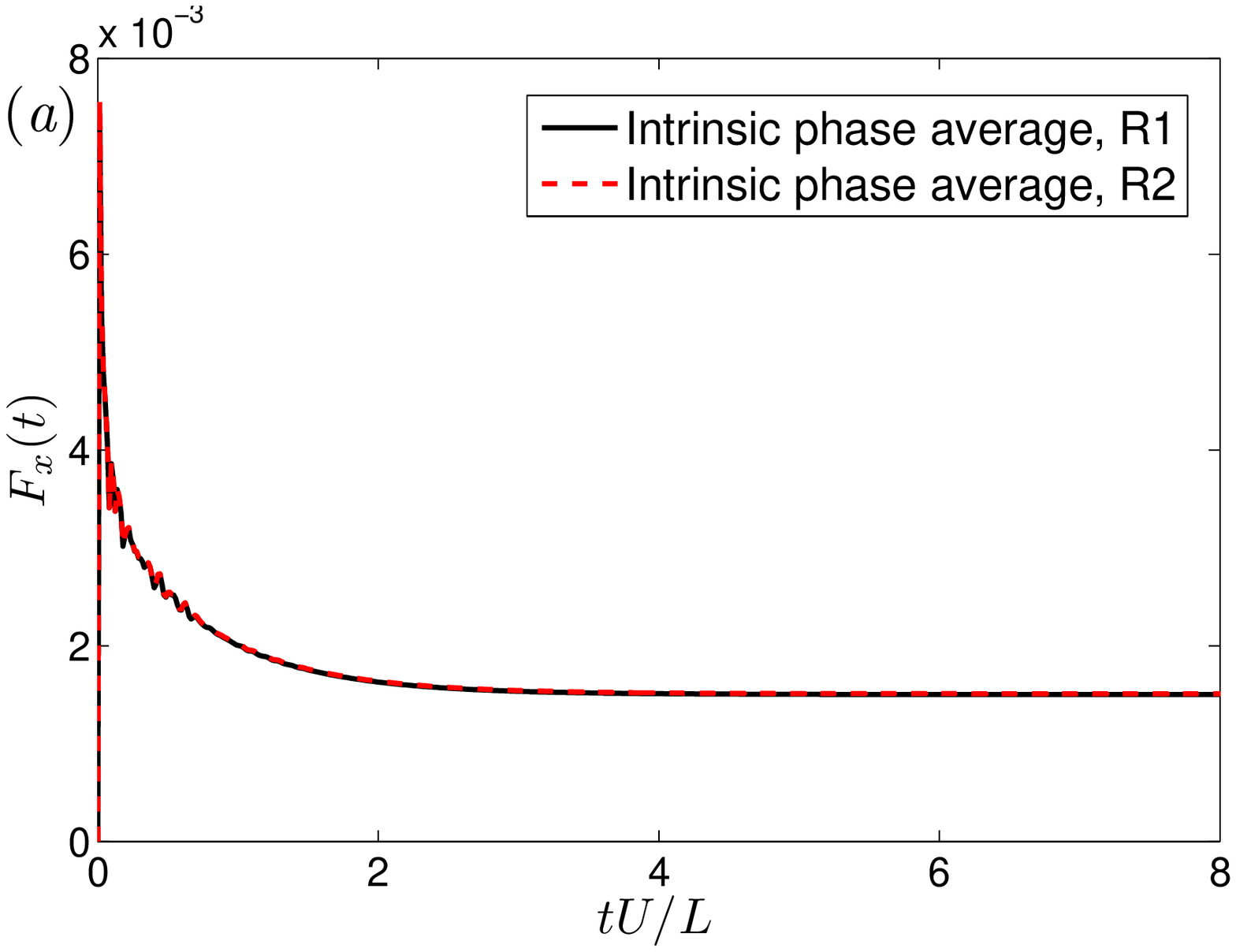}&
\includegraphics[width=0.49\textwidth,height=0.3\textheight]{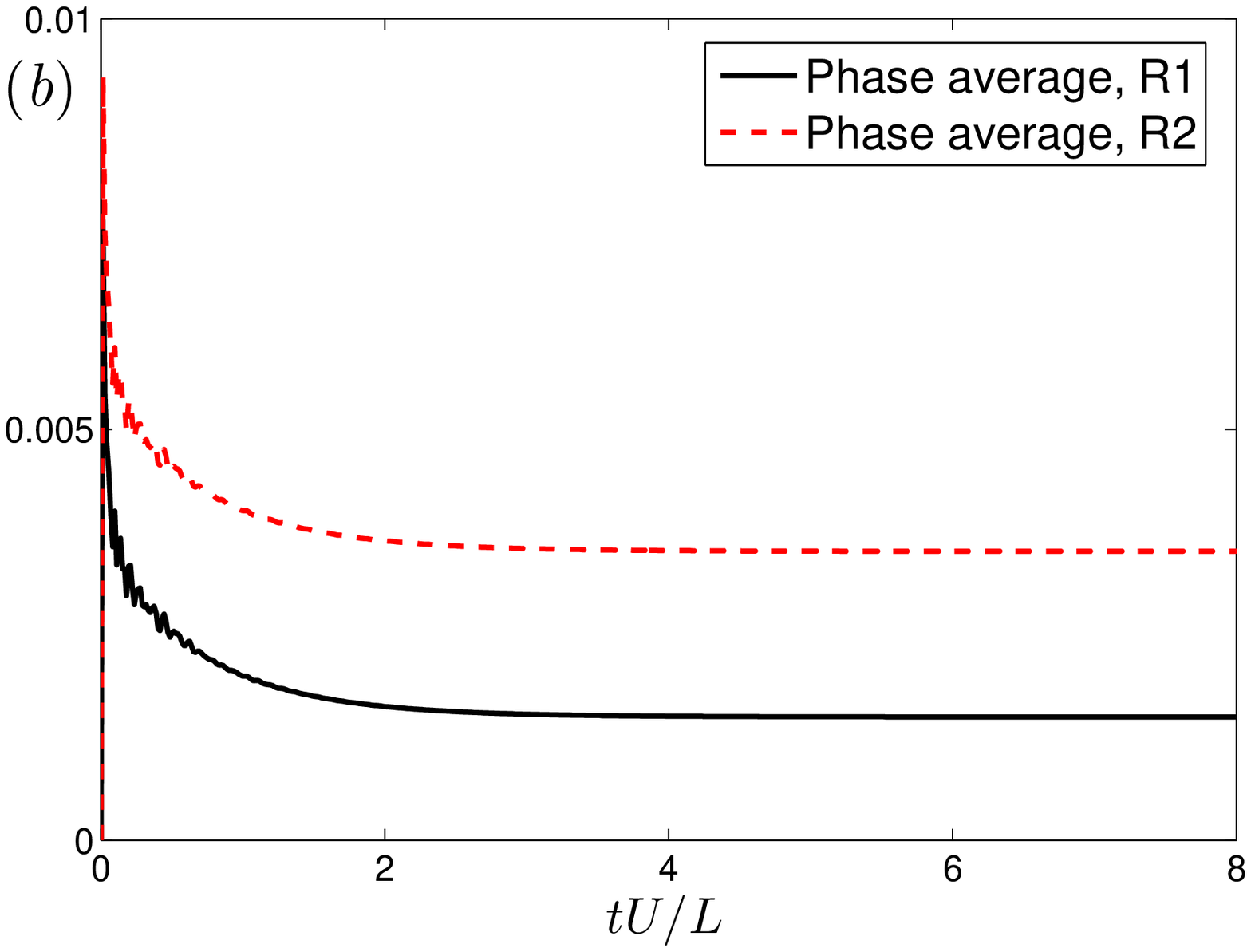}\\
\end{tabular}
\caption{The drag force $F_x$ on the porous cylinder as a function of time at $\varepsilon=0.7$. $F_x$ are both measured by (\textit{a}) the intrinsic phase average velocity, and (\textit{b}) the phase average velocity. Solid lines are the results in the frame of R1, and dashed lines are the results in the frame of R2.}
\label{fig:DragF1}
\end{figure}
Figure \ref{fig:DragF1} shows the time history of drag force on the porous cylinder, which are obtained respectively with the intrinsic phase average velocity ( Fig. \ref{fig:DragF1}$(\textit{a})$ ) and the phase average velocity ( Fig. \ref{fig:DragF1}$(\textit{b})$ ). As clearly seen from the figure, the computed drag forces by the intrinsic phase average velocity are in agreement with each other in the frames of R1 and R2, which satisfies the Galilean invariance of macroscopic equations. In contrast to this, the computed drag forces by the phase average velocity exhibit an obvious phase difference between the two frames, which lacks the Galilean invariance. These evidences confirm that only the macroscopic equations with intrinsic phase average velocity can preserve Galilean invariance.
\par
\begin{figure}
\begin{tabular}{cc}
\includegraphics[width=0.49\textwidth,height=0.3\textheight]{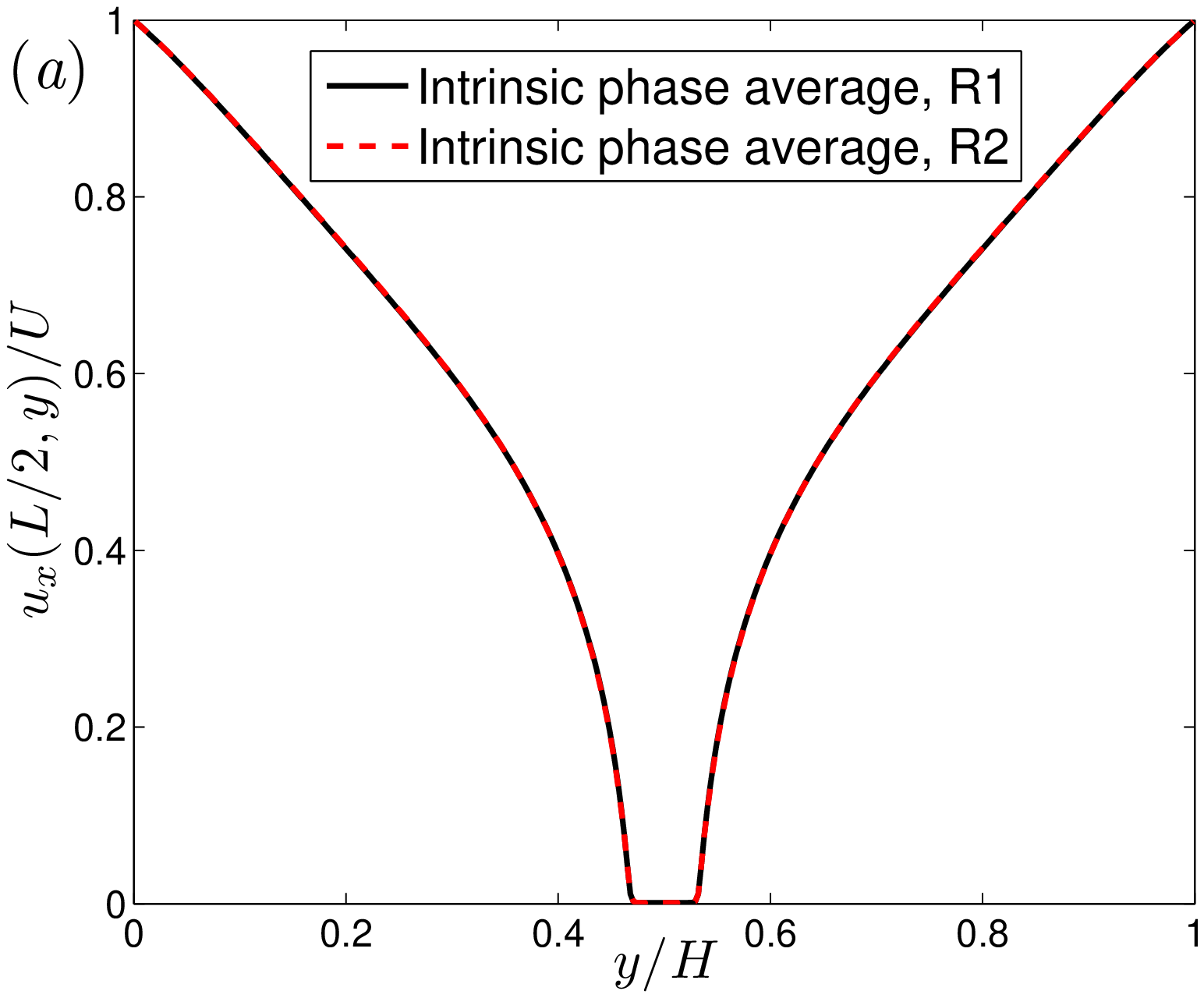}&
\includegraphics[width=0.49\textwidth,height=0.3\textheight]{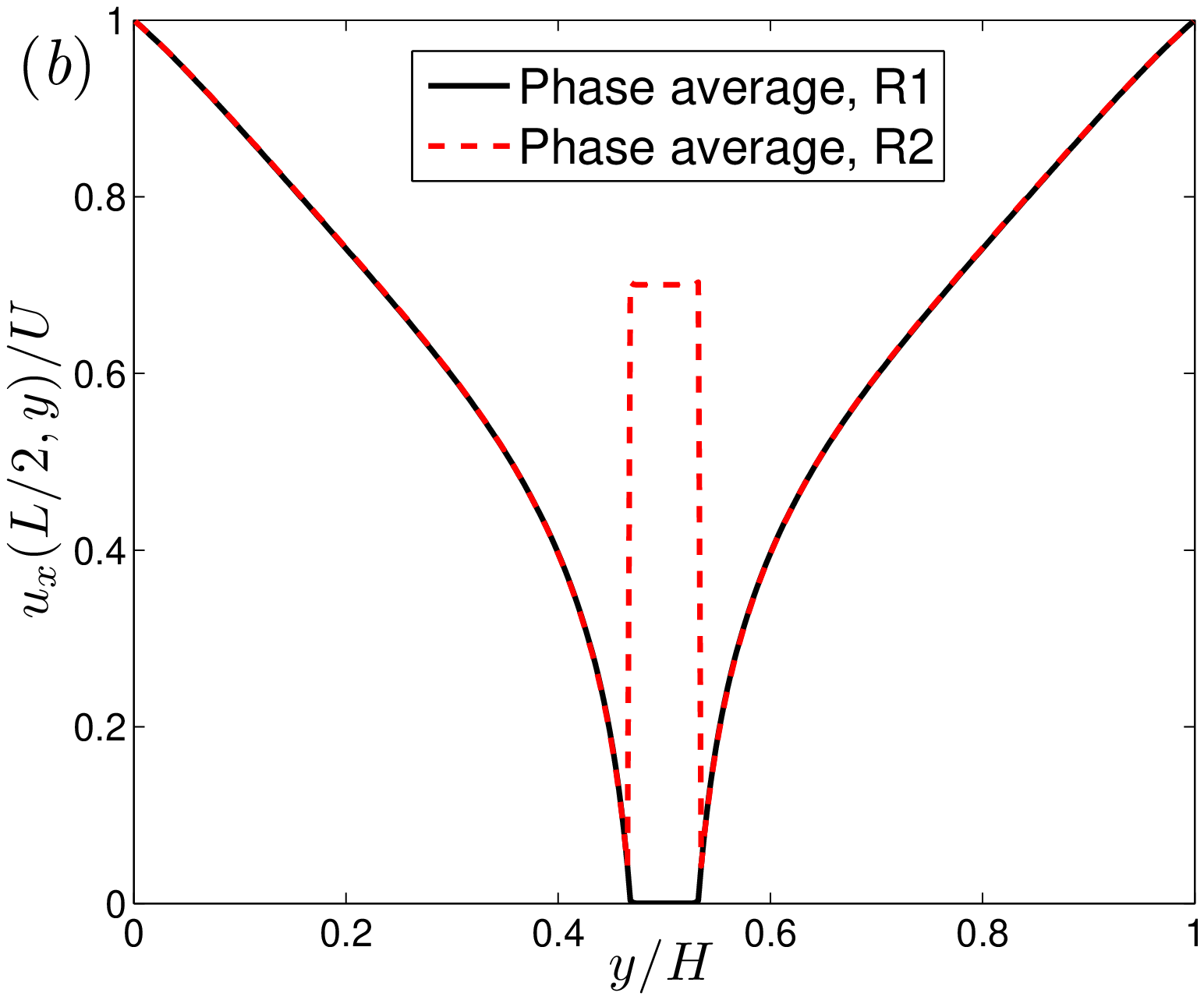}\\
\end{tabular}
\caption{Same as Fig. \ref{Galifig1} but $\epsilon=0.3$. (\textit{a}) The intrinsic phase average velocity. (\textit{b}) The phase average velocity.}
\label{Galifig2}
\end{figure}
\begin{figure}
\begin{tabular}{cc}
\includegraphics[width=0.52\textwidth,height=0.3\textheight]{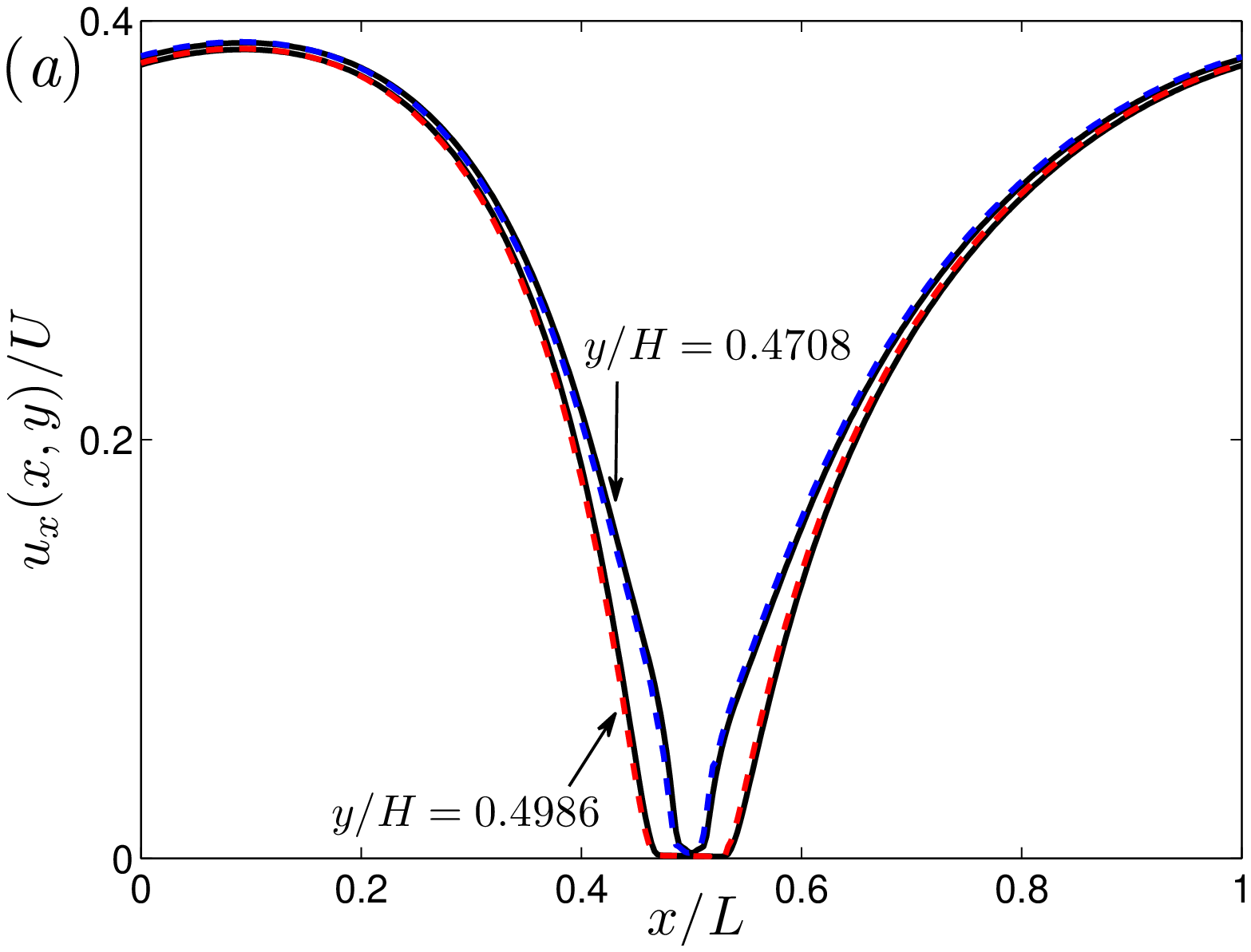}&
\includegraphics[width=0.52\textwidth,height=0.3\textheight]{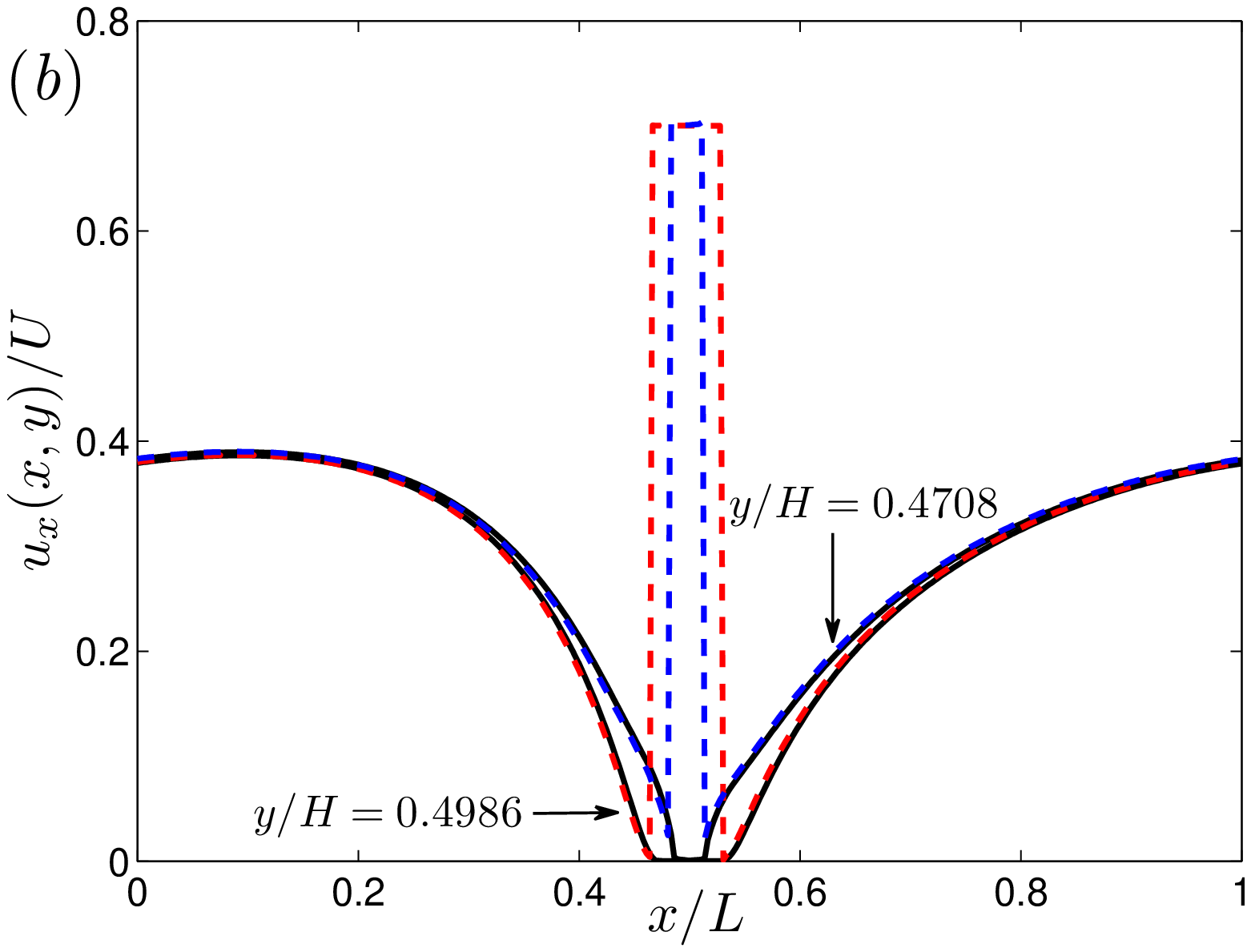}\\
\end{tabular}
\caption{Same as Fig. \ref{fig:Velcomp1} but $\varepsilon=0.3$. (\textit{a}) The intrinsic phase average velocity. (\textit{b}) The phase average velocity. Solid lines represent the results in the frame of R1, and dashed lines represent the results in the frame of R2.}
\label{fig:Velcomp2}
\end{figure}
\begin{figure}
\begin{tabular}{cc}
\includegraphics[width=0.49\textwidth,height=0.308\textheight]{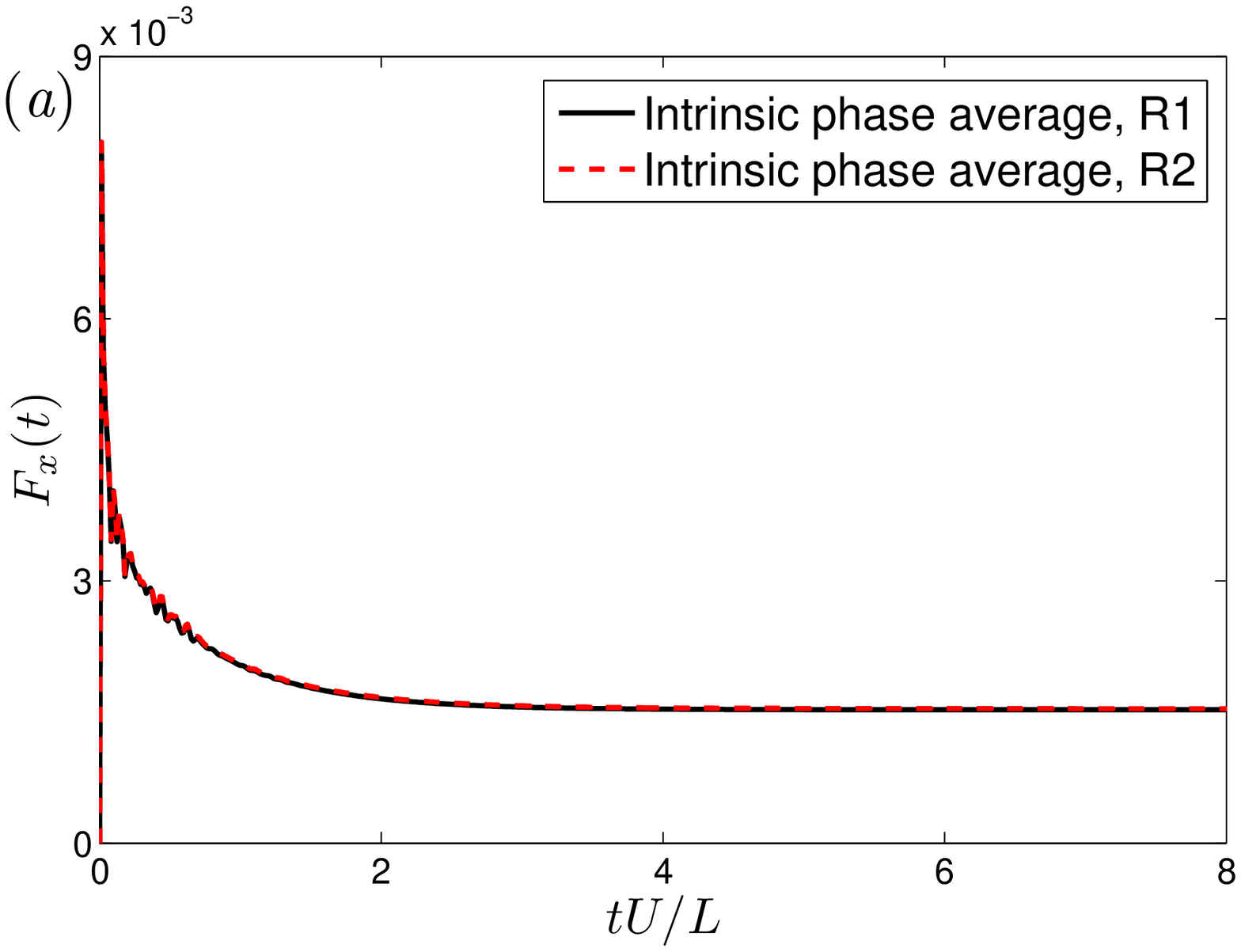}&
\includegraphics[width=0.49\textwidth,height=0.3\textheight]{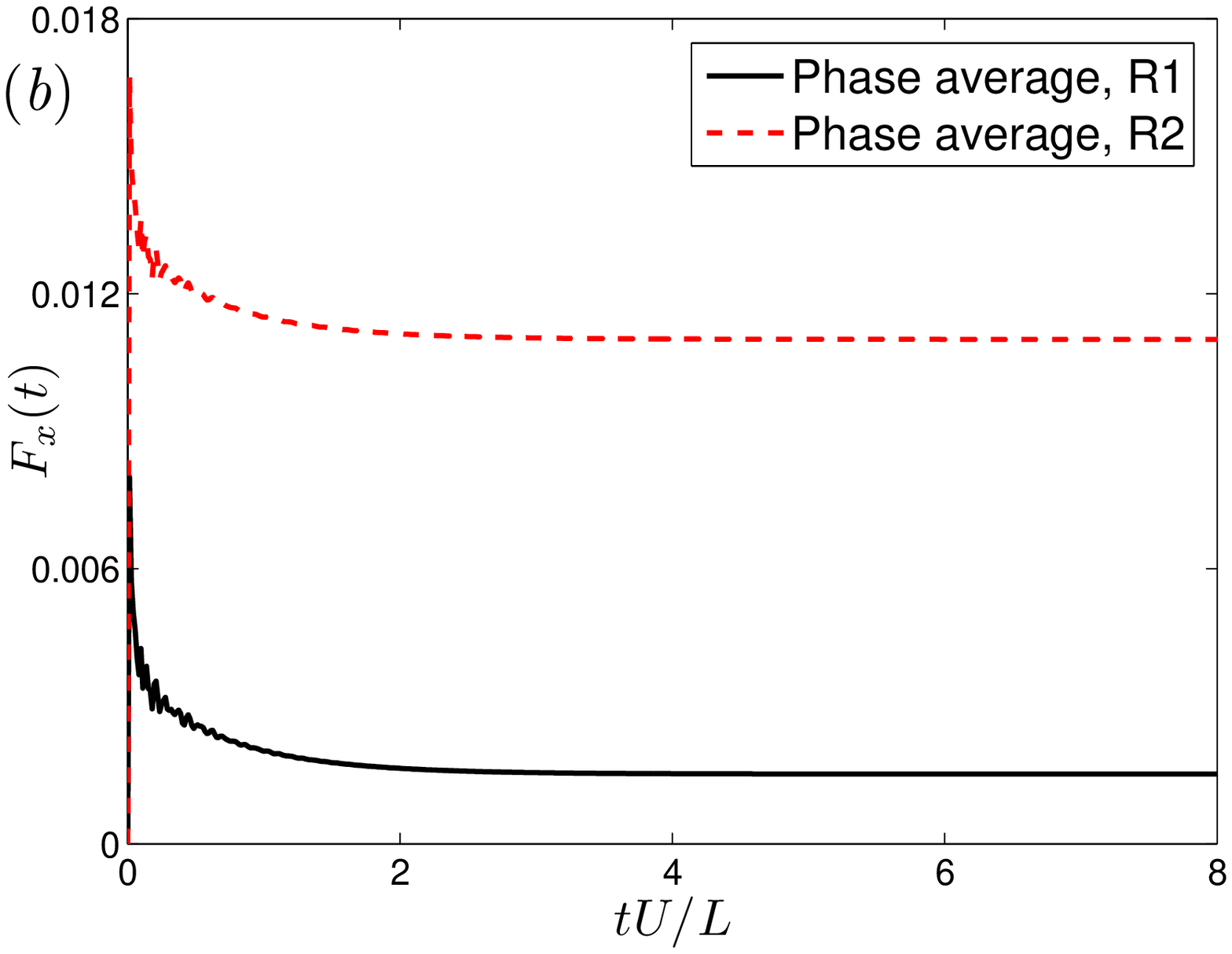}\\
\end{tabular}
\caption{Same as Fig. \ref{fig:DragF1} but $\varepsilon=0.3$. The drag force $F_x$ on the porous cylinder as a function of time at $\varepsilon=0.3$. $F_x$ are both measured by (\textit{a}) the intrinsic phase average velocity, and (\textit{b}) the phase average velocity.}
\label{fig:DragF2}
\end{figure}

The effect of porosity on the Galilean invariance are also investigated for the two volume-averaged velocities. For this purpose, simulations are carried out again in the two frames of reference while $\varepsilon$ is changed to $0.3$. The profiles of the two volume-averaged velocities are plotted in Figs. \ref{Galifig2} and \ref{fig:Velcomp2}.
Once again, results based on the intrinsic phase average velocity are in good agreement in the frames of R1 and R2, and significant differences between the two frames occur when the phase average velocity is used. Similar results are evident for the drag force on the cylinder, which is shown in Fig. \ref{fig:DragF2}.
This further demonstrates that Galilean invariance can only be satisfied with the intrinsic phase average velocity. Moreover, one can see that, as $\varepsilon$ decreases from $0.7$ to $0.3$, quantitative difference of the computed results with the phase average velocity between the two frames are more significant. Therefore, these facts confirm that the intrinsic phase average velocity instead of the phase average velocity should be identified as the superficial velocity in the derived macroscopic equations for flows with a moving porous body.

\section{Conclusions}
\label{Sec: conclusion}
In this paper, mass and momentum conservation equations for the fluid flows in a moving porous medium are derived by the technique of volume averaging of the microscopic equations at the pore sacle. Different from Darcy's law and the Brinkman equation, which are the two main models used in the literature, the resultant momentum equation includes the time derivative and nonlinear inertia terms. Nevertheless, the present momentum equation can reduce to these two models under creeping flow conditions. Furthermore, the developed macroscopic equations are valid for pure fluid and porous regions in the entire domain. In view of these results, the derived macroscopic equations can be regarded as the general governing equations. Another major contribution of this work is to investigate and testify, for the first time, the choice of the intrinsic phase average velocity as the superficial velocity for the porous particulate flow systems.

The LBE method is used to solve the macroscopic conservation equations, and the flow past a porous cylinder is simulated. By comparing the computed results in two frames of reference, Galilean invariance of the derived macroscopic equations in terms of the phase average velocity and the intrinsic phase average velocity are investigated. It is found that the intrinsic phase average velocities in the case of moving cylinder agree perfectly with those in the case of stationary cylinder, while the phase average velocities obtained in these two cases are not consistent with each other. Similar results are also obtained for the drag force on the porous cylinder. Furthermore, for the phase average velocity, the difference between the two frames of reference increases as the porosity of the cylinder decreases. This demonstrates that only with the use of the intrinsic phase average velocity, the derived macroscopic conservation equations are Galilean invariant.

Finally, the present work focuses on the macroscopic governing equations for the fuid flows in moving porous media. For the problem of heat transfer, the energy conservation equation in a moving porous medium could be considered. This will be left for future studies.

\section*{Acknowledgements}
{
  This work is financially supported by the National Natural Science Foundation of China under Grant Nos. 51125024 and 51390494, and by the Fundamental Research Funds for the Central Universities under Grant No. 2014TS119. LPW acknowledges support from the US National Science Foundation through CBET-1235974.
}

\appendix
\section{the volume-averaged equations in terms of phase average velocity}\label{AppSpa}
Applying relations $\langle\bm{u}_f\rangle=\varepsilon\langle\bm{u}_f\rangle^f$ and $\langle p_f\rangle=\varepsilon\langle p_f\rangle^f$ in Eqs. \eqref{Eq:MacroMass}, \eqref{Eq:IntrPhaseAvR} and \eqref{IntrPhaForce}, the following incompressible
macroscopic equations in terms of phase average velocity can be obtained
\begin{subequations}\label{Eq:PhavGove}
 \begin{equation}\label{Eq:MacroMassPhav}
   \nabla\cdot \langle\bm{u}_f\rangle =0,
 \end{equation}
 \begin{align}\label{Eq:PhavMoment}
   \rho_f\left[\frac{\partial \langle\bm{u}_f\rangle }{\partial t}+\nabla\cdot\left(\frac{\langle\bm{u}_f\rangle\langle\bm{u}_f\rangle}{\varepsilon}\right)\right]
    =-\nabla \langle p_f\rangle +\mu\nabla^2\langle\bm{u}_f\rangle+\bm{F},
\end{align}
where
\begin{equation}\label{Eq:PhavForce}
  \bm{F}=-\frac{\varepsilon\mu}{K}\left(\langle\bm{u}_f\rangle-\varepsilon\bm{V}_p\right)
   -\rho_f\frac{\varepsilon F_\varepsilon}{\sqrt{K}}\left(\langle\bm{u}_f\rangle-\varepsilon\bm{V}_p\right)
   \left|\langle\bm{u}_f\rangle-\varepsilon\bm{V}_p\right|+\varepsilon\rho_f\bm{g}.
\end{equation}
\end{subequations}

Correspondingly, the dimensionless equations are
\begin{subequations}\label{Eq:PhavGoveDimless}
 \begin{equation}\label{Eq:MacroMassPhavDimless}
   \nabla\cdot \langle\bm{u}_f\rangle =0,
 \end{equation}
 \begin{align}\label{Eq:PhavMomentDimless}
  \frac{\partial \langle\bm{u}_f\rangle }{\partial t}+\nabla\cdot\left(\frac{\langle\bm{u}_f\rangle\langle\bm{u}_f\rangle}{\varepsilon}\right)
    =-\nabla \langle p_f\rangle +\frac{1}{Re}\nabla^2\langle\bm{u}_f\rangle+\bm{F},
\end{align}
where
\begin{equation}\label{Eq:PhavForceDimless}
  \bm{F}=- \frac{\varepsilon}{ReDa}\left(\langle\bm{u}_f\rangle-\varepsilon\bm{V}_p\right)
   -\frac{\varepsilon F_\varepsilon}{\sqrt{Da}}\left(\langle\bm{u}_f\rangle-\varepsilon\bm{V}_p\right)
   \left|\langle\bm{u}_f\rangle-\varepsilon\bm{V}_p\right|+\varepsilon\bm{g}.
\end{equation}
\end{subequations}

\section{Lattice Boltzmann equation model for the volume-averaged macroscopic equations}
\label{Appen:MRTMod}
In this appendix, a LBE model is provided to solve the derived volume-averaged equations for the flows in a moving porous medium.
\subsection{The LBE model for the intrinsic phase average equations}
Eq. \eqref{Eq:InPhavGove} is first considered, and $\bm{u}$ is temporarily used to denote the intrinsic phase average velocity $\langle\bm{u}_f\rangle^f$. The LBE is written as
\begin{equation}\label{Eq-porousMRT}
    f_i(\bm{x}+\bm{c}_i\delta_t,t+\delta_t)-f_i(\bm{x},t)=-S_{i\alpha}[~f_\alpha(\bm{x},t)- f_\alpha^{(eq)}(\bm{x},t)~]
            +\delta_t\bigl(\delta_{i\alpha}-\frac{{S}_{i\alpha}}{2}\bigl)J_\alpha(\bm{x},t),
\end{equation}
where $f_i(\bm{x},t)$ is the discrete distribution function for particles moving with velocity $\bm{c}_i$ at point $\bm{x}$ at time $t$, $S_{i\alpha}$ is the element of collision matrix $\bm{S}$, $\delta_t$ is the time step, $\delta_{i\alpha}$ stands for Kronecker delta function, the equilibrium distribution function $f_\alpha^{(eq)}$ and forcing term $J_\alpha$ are respectively given by
\begin{equation}
  f_\alpha^{(eq)}=\omega_\alpha\rho\biggl[1+\frac{\bm{c}_\alpha\cdot\bm{u}}{c_s^2}
  +\frac{\bm{uu}:(\bm{c}_\alpha\bm{c}_\alpha-c_s^2\bm{I})}{2c_s^4}\biggr],
\end{equation}
\begin{equation}
   J_\alpha=\omega_\alpha\rho\biggl[\frac{\bm{c}_\alpha\cdot\bm{F}}{c_s^2}
   +\frac{\bm{uF}:(\bm{c}_\alpha\bm{c}_\alpha-c_s^2\bm{I})}{c_s^4}\biggr],
\end{equation}
where $\omega_\alpha$ is the weight coefficient related to the discrete velocity set, $c_s$ is the sound speed, and $\bm{F}$ is the total body force given by Eq. \eqref{Eq:InPhavForce}.
\par
For 2D flow problems, the discrete velocity set $\{\bm{c}_i\}$ in the above LBE model is given by
\begin{equation}\label{disvelocity}
 \bm{c}_i=
  \begin{cases}(\ 0,0),&i=0,\\
  \bigl(\cos\bigl[(i-1)\pi/2\bigr],\sin\bigl[(i-1)\pi/2\bigr]\bigr)\ c,&i=1-4,\\
   \bigl(\cos\bigl[(i-1)\pi/2+\pi/4\bigr],\sin\bigl[(i-1)\pi/2+\pi/4\bigr]\bigr)\sqrt{2}\ c,&i=5-8,
  \end{cases}
\end{equation}
where $c=\delta_x/\delta_t$ is the lattice speed with $\delta_x$ being the lattice spacing, $c_s=c/\sqrt{3}$ is the sound speed. The weight coefficients are given by $\omega_0=4/9$, $\omega_{1-4}=1/9$, $\omega_{5-8}=1/36$.
\par
The evolution equation, Eq. \eqref{Eq-porousMRT} for the distribution function is decomposed into two substeps:
the collision and propagation substeps. Via a linear transformation matrix $\bm{M}$ at $c=1$, the collision step is executed in the moment space \cite{Ddhum92, Lalleme&Luo00},
\begin{equation}\label{MRT-col}
  \bm{m}^+=\bm{m}-\widetilde{\bm{S}}\bigl[\bm{m}-\bm{m}^{(eq)}\bigr]+\delta_t\biggl(\bm{I}-\frac{\widetilde{\bm{S}}}{2}\biggl)~\hat{\bm{J}}\; ,
\end{equation}
where $\bm{m}$ and $\bm{m}^+$ are the moment and its post-collision moment vectors,
\begin{align*}
   \bm{m}=\bm{M}\cdot\bm{f}=(m_0,m_1,\cdots,m_8)^T,\quad
    \bm{f}=\bm{M}^{-1}\cdot\bm{m}=(f_0,f_1,\cdots,f_8)^T,
\end{align*}
and $\widetilde{\bm{S}}=\bm{M}\bm{S}\bm{M}^{-1}$ is diagonal in the moment space,
\begin{equation}
   \widetilde{\bm{S}}={diag}(s_0,s_1,\ldots,s_8),\quad s_i \geqslant 0
\end{equation}
where $s_i$~($i=0,1,\cdots,8$) is the relaxation rate for moment vector $m_i$.

The equilibrium moment $\bm{m}^{(eq)}=\bm{M}\bm{f}^{(eq)}$ and the moments of forcing term $\hat{\bm{J}}=\bm{M}\bm{J}$ can be computed,
\begin{alignat}{2}\label{InPhaMoment}
  \bm{m}^{(eq)}=
  {
  \left(
  \begin{array}{c}
    \rho\\
    \rho (-2+3u_x^2+3u_y^2)\\
    -\rho (-1+3u_x^2+3u_y^2)\\
    \rho u_x\\
    -\rho u_x\\
    \rho u_y\\
    -\rho u_y\\
    \rho (u_x^2-u_y^2)\\
    \rho u_xu_y\\
  \end{array}
\right), }
 &\qquad& \hat{\bm{J}}=
 {
  \left(
  \begin{array}{c}
    0\\
    6\rho (F_xu_x+F_yu_y)\\
    -6\rho (F_xu_x+F_yu_y)\\
    \rho F_x\\
    -\rho F_x\\
    \rho F_y\\
    -\rho F_y\\
    2\rho (F_xu_x-F_yu_y)\\
    \rho (F_xu_y+F_yu_x)\\
  \end{array}
\right), }
\end{alignat}
where $\bm{u}=(u_x, u_y)$ and $\bm{F}=(F_x, F_y)$ are the fluid velocity and the total body force, respectively.

By transforming $\bm{m}^+$ back to the  velocity space, the propagation process is implemented in velocity space,
\begin{equation}\label{MRT-str}
  \bm{f}^+ =\bm{M}^{-1}\bm{m}^+\;,\quad
   f_i(\bm{x}+\bm{c}_i\delta_t,t+\delta_t)=f_i^+(\bm{x},t)~.
\end{equation}
\par
The macroscopic properties are related to the distribution functions by
\begin{equation}\label{Pormacro}
   \rho=\sum_if_i~,\qquad \rho\bm{u}=\sum_i\bm{c}_if_i+\frac{\delta_t}{2}\rho\bm{F}~.
\end{equation}
\par
Through solving a quadratic equation, $\bm{u}$ is computed by \cite{Guo&Zh02},
\begin{equation}\label{InPhaEqucom1}
   \bm{u}=\frac{\bm{v}}{d_0+\sqrt{d_0^2+d_1|\bm{v}|}}+\bm{V}_p~,
\end{equation}
where $\bm{v}$ is an temporary velocity and defined as
\begin{equation}\label{InPhaEqucom2}
   \rho\bm{v}=\sum_i\bm{c}_if_i-\rho\bm{V}_p+\frac{\delta_t}{2}\rho\bm{g}~.
\end{equation}
The two parameters $d_0$ and $d_1$ are given by
\begin{equation}\label{InPhaEqucom3}
   d_0=\frac{1}{2}\left(1+\frac{\delta_t}{2}\frac{\varepsilon\nu}{K}\right), \qquad d_1=\frac{\delta_t}{2}\frac{\varepsilon^2 F_\epsilon}{\sqrt{K}}~.
\end{equation}

From the above LBE model, the following equations can be obtained through the multiscale analysis:
\begin{subequations}\label{MacroMRT}
  \begin{gather}
    \frac{\partial \rho}{\partial t}+\nabla\cdot(\rho\bm{u})=0,\\
    \frac{\partial(\rho\bm{u})}{\partial t}+\nabla\cdot\left(\rho\bm{uu}\right)=-\nabla p + \nabla\cdot\bm{\varrho}+\rho\bm{F},
  \end{gather}
\end{subequations}
where $p=c_s^2\rho$ is the pressure, the shear stress of fluid $\bm{\varrho}$ is given by
$$
   \bm{\varrho}=\rho\zeta(\bm{\nabla}\cdot\bm{u})\bm{I}+2\rho\nu\overset{\circ}{\bm{S}},
   \qquad\overset{\circ}{\bm{S}}=\bm{S}-\frac{1}{2}\text{Tr}(\bm{S})\bm{I},
$$
where $\bm{S}$ is the shear rate defined by $\bm{S}=(S_{ij})=(\partial_iu_j+\partial_ju_i)/2$, ``Tr'' is the trace operator, and $\nu$ and $\zeta$ are the kinematic viscosity and bulk viscosity of the fluid, respectively. For the incompressible flow,
it is noted that Eq. \eqref{MacroMRT} turns into Eq. \eqref{Eq:InPhavGove}.

In the multiscale analysis for the LBE model, it is required that $s_7=s_8$ and $s_4=s_6$. At the same time, the relaxation rates $s_1$ and $s_7$ are determined respectively by $\zeta$ and $\nu$
\begin{align}\label{Visequ}
  \zeta=\frac{1}{3}\bigl(\frac{1}{s_1}-\frac{1}{2}\bigr)\delta_t,\qquad
  \nu=\frac{1}{3}\bigl(\frac{1}{s_7}-\frac{1}{2}\bigr)\delta_t~.
\end{align}

From a numerical point of view, the process for solving the dimensionless equation, Eq. \eqref{Eq:InPhavGoveDimless}, by the LBE model is exactly the same as Eq. \eqref{Eq:InPhavGove}, in which the parameters $d_0$ and $d_1$ are renewed to the following formula:
\begin{equation}\label{varicomput}
   d_0=\frac{1}{2}\left(1+\frac{\delta_t}{2}\frac{\varepsilon}{ReDa}\right), \qquad d_1=\frac{\delta_t}{2}\frac{\varepsilon^2 F_\epsilon}{\sqrt{Da}}~.
\end{equation}

\subsection{The LBE model for the phase average equations}
As for numerical solution of the phase-averaged equations, Eq. \eqref{Eq:PhavGove}, the above LBE model is also employed except for the following changements:
\par
(1) the moment vectors $\bm{m}^{(eq)}$ and $\hat{\bm{J}}$ in Eq. \eqref{InPhaMoment} are renewed as
\begin{alignat}{2}
  \bm{m}^{(eq)}=
  {
  \left(
  \begin{array}{c}
    \varepsilon\rho\\
    \rho (-2\varepsilon+3u_x^2/\varepsilon+3u_y^2/\varepsilon)\\
    -\rho (-\varepsilon+3u_x^2/\varepsilon+3u_y^2/\varepsilon)\\
    \rho u_x\\
    -\rho u_x\\
    \rho u_y\\
    -\rho u_y\\
    \rho (u_x^2-u_y^2)/\varepsilon\\
    \rho u_xu_y/\varepsilon\\
  \end{array}
\right), }
 &\qquad& \hat{\bm{J}}=
 {
  \left(
  \begin{array}{c}
    0\\
    6\rho (F_xu_x+F_yu_y)/\varepsilon\\
    -6\rho (F_xu_x+F_yu_y)/\varepsilon\\
    \rho F_x\\
    -\rho F_x\\
    \rho F_y\\
    -\rho F_y\\
    2\rho (F_xu_x-F_yu_y)/\varepsilon\\
    \rho (F_xu_y+F_yu_x)/\varepsilon\\
  \end{array}
\right). }
\end{alignat}
\par
(2) The fluid velocity $\bm{u}=(u_x, u_y)$ takes the phase average form, and the total body force $\bm{F}=(F_x, F_y)$ follows Eq. \eqref{Eq:PhavForce}.
\par
(3) The formulae of Eq. \eqref{Pormacro} are modified to compute the fluid density and velocity,
\begin{equation}\label{PhaPormacro}
   \varepsilon\rho=\sum_if_i~,\qquad \rho\bm{u}=\sum_i\bm{c}_if_i+\frac{\delta_t}{2}\rho\bm{F}~.
\end{equation}
\par
(4) Eqs. \eqref{InPhaEqucom1}~-~\eqref{InPhaEqucom3} are updated as shown below:
\begin{equation}
   \bm{u}=\frac{\bm{v}}{d_0+\sqrt{d_0^2+d_1|\bm{v}|}}+\varepsilon\bm{V}_p~,
\end{equation}
\begin{equation}
   \rho\bm{v}=\sum_i\bm{c}_if_i-\rho\varepsilon\bm{V}_p+\frac{\delta_t}{2}\rho\varepsilon\bm{g}~,
\end{equation}
\begin{equation}\label{varicomput}
   d_0=\frac{1}{2}\left(1+\frac{\delta_t}{2}\frac{\varepsilon\nu}{K}\right)~, \qquad d_1=\frac{\delta_t}{2}\frac{\varepsilon F_\epsilon}{\sqrt{K}}~.
\end{equation}
\par
To solve the dimensionless equations, Eq. \eqref{Eq:PhavGoveDimless}, the LBE model for Eq. \eqref{Eq:PhavGove} is also used while employing the following new parameters:
\begin{equation}\label{varicomputDimless}
   d_0=\frac{1}{2}\left(1+\frac{\delta_t}{2}\frac{\varepsilon}{ReDa}\right)~, \qquad d_1=\frac{\delta_t}{2}\frac{\varepsilon F_\epsilon}{\sqrt{Da}}~.
\end{equation}





\end{document}